\documentclass[apj]{emulateapj}
\usepackage{color}
\usepackage{latexsym, graphicx, amssymb, longtable,epsf}

\long\def\symbolfootnote[#1]#2{\begingroup\def\thefootnote{\fnsymbol{footnote}}
\footnote[#1]{#2}\endgroup}

\shorttitle{Planetesimals Pile Up In Inclined Binary Systems}
\shortauthors{Xie et al.}

\begin{document}

\title{Planet Formation In Highly Inclined Binary Systems \\ I. Planetesimals Jump Inwards And Pile Up}

\author{Ji-Wei Xie$^{1,2}$, Matthew J. Payne$^2$, Philippe Th{\'e}bault$^3$, Ji-Lin Zhou$^1$,  Jian Ge$^2$, }
\affil{$^1$Department of Astronomy \& Key Laboratory of Modern Astronomy and Astrophysics in Ministry of Education, Nanjing University, 210093, China}
\affil{$^2$Department of Astronomy, University of Florida, Gainesville, FL, 32611-2055, USA}
\affil{$^3$Observatoire de Paris, Section de Meudon, F-92195 Meudon Principal Cedex, France}
\email{xiejiwei@gmail.com}

\begin{abstract}
Most detected planet-bearing binaries are in wide orbits, for which a high inclination, $i_B$, between the binary orbital plane and the plane of the planetary disk around the primary is likely to be common. In this paper, we investigate the intermediate stages - from planetesimals to planetary embryos/cores - of planet formation in such highly inclined cases. Our focus is on the effects of gas drag on the planetesimals' orbital evolution, in particular on the evolution of the planetesimals' semimajor axis distribution and their mutual relative velocities. We first demonstrate that a non-evolving axisymmetric disk model is a good approximation for studying the effects of gas drag on a planetesimal in the highly inclined case ($30^\circ <i_B<150^\circ$). We then find that gas drag plays a crucial role, and the results can be generally divided into two categories, i.e., the Kozai-on regime and the Kozai-off regime, depending on the specific value of $i_B$. For both regimes, a robust outcome over a wide range of parameters is that, planetesimals migrate/jump inwards and pile up, leading to a severely truncated and dense planetesimal disk around the primary. In this compact and dense disk, collision rates are high but relative velocities are low, providing conditions which are favorable for planetesimal growth, and potentially allow for the subsequent formation of planets.
\end{abstract}

\keywords{methods: numerical --- planets and satellites: formation }
%

\section{Introduction}
Planets in binaries could be common as stars are usually born in
binary or multi-stars systems \citep{DM 91, Mat 00, Duc 07}. Although current observations show that the multiplicity-rate of the detected exoplanet host stars is around $17\% $ \citep{MN 09}, this fraction should be considered as a lower limit because of noticeable selection effects against binaries in planet searches \citep{Egg 10}.  
Most of the currently known
planet-bearing binary systems are wide S-types \citep{Egg 04, DB 07, Hag
10}, meaning the companion star acts as a distant satellite,
typically orbiting the inner star-planet system over 100 AU away.
Nevertheless, there are currently four systems with smaller
separation of $\sim 20$ AU, including the $\gamma$ Cephei \citep{Hat
03}, GJ 86, \citep{Que 00}, HD 41004 \citep{Zuc 04}, and HD 196885
\citep{Cor 08, Cha 10}. 
In addition to the planets in circumprimary (S-type) orbits discussed above, planets in circumbinary (P-type) orbits have been found in two systems \citep{Sig 03, Lee 09}. While it is possible that the recent detections of numerous short period eclipsing binaries by the Kepler mission \citep{Prs 11, Sla 11} may promote additional such circumbinary detections, their short periods are likely to preclude their use in the characterization of the circumprimary planets on which we concentrate in this investigation.

The properties and diversities of planet-bearing double-star systems provide additional challenges and constraints for theories of planet formation.
Current theories of the core-accretion model of planet formation in single-planet systems \citep{Lis 93, Arm 10} envisage formation proceeding via a number of successive stages.
In the first step, sub-mm dust grains are converted into
0.1-1000 km-sized planetesimals, the specific mechanism for which may
involve (i) some form of grain coagulation \citep{WC 93, Wei 97, Joh 08, TW 09} and/or (ii) some form of fragmentation due to instabilities
in the solid sub-component of the protoplanetary gas disk \citep{Saf 69, GW 73, YS 02, YG 05, YJ 07, Cuz 08}. Reviews
of these topics are given by \citet{BW 08} and \citet{CY 10}.
In the second stage, mutual (planetesimal-planetesimal) collisions
commence and the planetesimals grow via runaway and oligarchic
phases \citep{WS 89, KI 96, KI 98, Raf 03, Raf 04} to form protoplanetary embryos.
Depending on the birth-rate and birth size of planetesimals, there could be a transition phase$-$``snowball" phase$-$ between the first and second stages \citep{Xie 10b}.
In the third stage, giant impacts between these embryos commence,
leading to the formation of full-size terrestrial planets and
gas-giant cores \citep{CW 98, LA 03, Kok 06}.

Perturbative effects from stellar companions render the
planet formation process in binary systems even more complex than
that in single-star systems. For the first stage of
planet formation outlined above, it is still unclear whether and/or how the formation of planetesimals from dust is affected by the binary star \citep{Pas 08, Zso 10}; whereas for the third stage, it is thought that planet formation is probably
unaffected or even hastened \citep{Bar 02,
Qui 04, QL 06, Qui 07, Gue 08} in a region that extends to the outer limit for orbital stability, i.e., out to $\sim$3-5 AU for a typical tight binary of semi-major axis 20 AU \citep{HW 99}.

However, the
intermediate (second)
stage of the formation process is much more
problematic, as the outcome of planetesimal-planetesimal
collisions is highly sensitive to the relative velocities between
the bodies during any encounter \citep{BA 99, SL 09}. The perturbations from
a binary companion can excite planetesimal orbits and increase their mutual impact velocity, $\Delta V$, to values that might exceed their escape velocity or even the critical velocity for the onset of eroding collisions \citep{Hep 78,
Whi 98}. 

Some earlier investigations \citep{MS 00} suggested that
the combination of binary perturbations and gas damping could force
a strong pericenter alignment between planetesimal orbits, reducing
$\Delta V$. However the work of  \citet{The 06}
showed that it is only similarly sized objects that experience this
reduction, as in general the preferred pericenter alignment is
strongly size-dependent, leading to an increasing
$\Delta V$ for objects of different sizes. Applying such an analysis
to the $\alpha$-Centauri system ($a_B\sim 18$ AU), \citet{The 08, The 09} showed that ``accretion-friendly" zones exist only for regions within $\sim 0.5 - 0.75$ AU of an individual star within the binary system. Furthermore, the situation would become much more unfavorable for planetesimal accretion if the eccentricity and precision of the gas disk are considered \citep{Par 08, Bea 10}.
Put simply, the planetesimals-to-embryos stage is strongly affected by binary effects for a wide range of parameter space in a circumprimary protoplanetary disk, and it is probably the main obstacle to the formation of planets in close binaries.
Some of the deleterious effects of
this gas drag could be mitigated if the effects of a dissipating gas
disk are taken into account \citep{XZ 08}, but the effect is
rather inefficient for small bodies and requires a long timespan
($\sim 10^6$ years) for the planetesimals to reach a low $\Delta V$
state. 
One possible solution to this critical problem has been investigated by \citet{PL 10} and \citet{Xie 10b}. Both
 have shown that dust accretion onto planetesimals may provide a safe way for bodies to grow through the problematic 1 to 50 km size range for which the perturbed environment of the binary can prevent mutual accretion of planetesimals. However, several problems still remain unclear, such as the planetesimal birth rate and the efficiency of dust accretion.

Much of the previously mentioned work had considered only systems in
which the binary and planetesimal disk were coplanar, so \citet{XZ 09} went on to consider the effects of small relative inclinations
($i_B< 10^{\circ}$) between the binary orbit and the plane of the
protoplanetary disk. Overall, they found that small relative
inclinations significantly favor accretion through a reduction in
$\Delta V$. Then \citet{Xie 10a} demonstrated that if the disk
has a small inclination, the accretion friendly zone can be
pushed out to $\sim 1-2$ AU for a system such as $\alpha$-Centauri.

\citet{Mar 09} investigated cases of much higher
inclination, a situation which may be particularly common amongst
intermediate separation binaries ($40 - 100$ AU), where one does not
expect there to be significant alignment between disk and binary
plane \citep{Hal 94, Jen 04, Mon 06, Mon 07}. In their investigation they
neglected gas drag, arguing that the planetesimal spends the
majority of it's orbit outside of the gas disk and hence drag
effects are negligible. However, we feel that this argument is
specious, as it ignores two facts which are crucial. One is that the planetesimals will cross the gas disk at high inclination, $\hat{i}$, and
very high relative velocities, $V_{rel}$, thus enhancing gas drag, $F_d$. This enhancement could be very significant despite the planetesimal spending less of its orbital time in the disk ($t_{in}$). This is because $F_d$ is proportional to the square of $\hat{i}$\symbolfootnote[2]{Taking the approximation for high $\hat{i}$ cases i.e., $V_{rel}\sim \hat{i}V_k$, where $V_k$ denotes the Keplerian velocity, we get $F_d\propto \hat{i}^2$ from Eqn.\ref{fd}.},
while $t_{in}$ is inversely proportional to the first power of $\hat{i}$. Thus, in total, the planetesimal experiences an increasing momentum change
 ($F_d*t_{in}$) with increasing $\hat{i}$ from gas drag.
The other pertinent fact is
that planetesimals in sufficiently inclined orbits
are likely to experience significant Kozai-driving \citep{Koz 62} from
the binary companion, and hence spend some significant portion of
their time with extremely high eccentricities and associated low
pericenters. Given some typical power-law density profile for a
proto-planetary gas disk \citep{Hay 81, Pri 81, IL 04}, this will mean that such planetesimals will be interacting
with an extremely dense region of gas at pericenter, and hence will
be likely to suffer significant drag even during their brief passage
through the gas disk.

Given the demonstrated importance of inclination and gas drag
effects in low inclination binary systems \citep{The 06, XZ 09}, we feel that a full
investigation of planetesimal evolution in highly inclined binary
systems requires an adequate treatment of gas damping. As such, we
perform N-body simulations of planetesimal dynamics in highly
inclined binary systems, adding drag forces to simulate the effects
of the planetesimals interacting with a gas disk.

We structure our investigation of the planetesimal dynamics in such
highly inclined systems as follows:
In \S\ref{SECN:MODEL} we investigate the effects of different gas disk
models (circular and eccentric, precessing and stationary) and justify our choice of the simplest gas disk model.
In \S\ref{SECN:SINGLE} we look at the behavior of individual
particles as they interact with the gas disk and inclined binary
companion, using both analytical and numerical methods.
Then in \S\ref{SECN:SWARM} we investigate the bulk properties
of a disk of planetesimals in such a highly inclined system, showing
the overall changes in surface density that can take place, and their dependence on parameters such as the binary configuration, the radial size of the planetesimals and the gas disk density.
Discussions of the implications
for planet formation and the limitations of our approach
are made in \S\ref{SECN:DISCUSS}.
Finally,  we summarize in \S\ref{SECN:SUMMARY}.

\section{Model}\label{SECN:MODEL}
\begin{figure}
\begin{center}
\includegraphics[width=0.8\textwidth,bb=0 0 600 800]{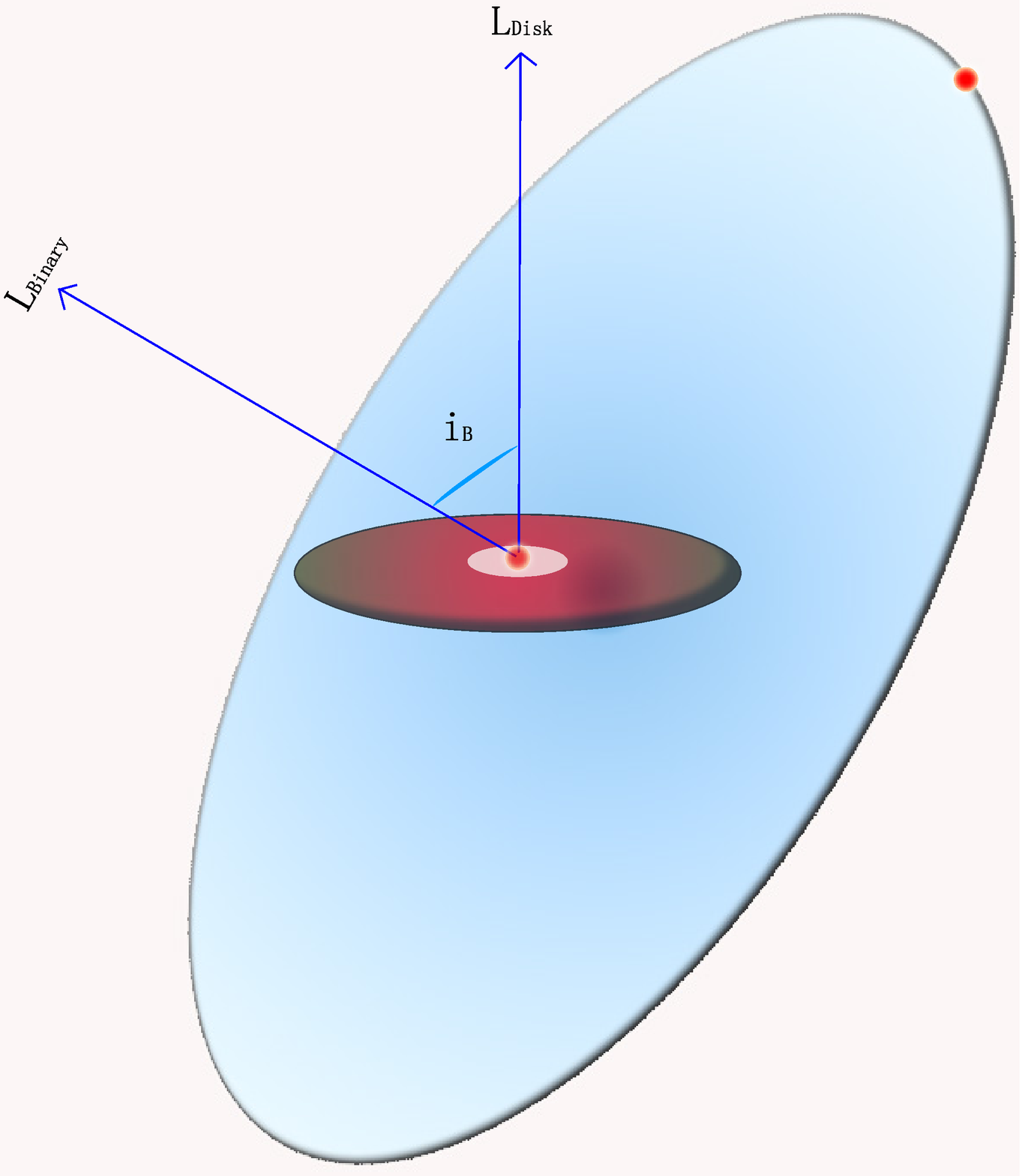}
  \caption{Sketch of the orbital configuration of the highly inclined binary system studied in this paper. $L_{Disk}$ and $L_{Binary}$ are the normals of the gas disk plane and the binary orbital plane, respectively. The angle $i_B$ between $L_{Disk}$ and $L_{Binary}$ corresponds to the mutual inclination of the two planes.
  }
  \label{orbit}
   \end{center}
\end{figure}

Fig.\ref{orbit} is a sketch of the orbital configuration of the binary stars and the circumprimary disk studied in this paper. The companion star acts as an outer perturber on an inclined orbit with mutual inclination of $i_B$ around the inner circumprimary disk.
Planetesimals are initially embedded in the gas disk and are always treated as test particles, i.e. their growth and mutual gravitational interactions are not included and their motions are affected only by the gravitational forces from the two stars and the effects of
the gas disk. The gas disk may have two effects on the motion of planetesimals.
One is the gravity of the gas disk, which is ignored in this paper and will be addressed in detail in our forthcoming paper (see the discussion in \S\,\ref{SECN:LIMITATION}). The other effect is the gas drag, which is generally applied to planetesimals whenever they have non-zero
relative velocities ($V_{rel}$) to the gas. Throughout this paper,
the disk gravity is ignored and we only focus on the effect of gas
drag to investigate the planetesimals' motions in binary star
systems of different orbital configurations. We use the Mercury code with the Bulirsch-Stoer integrator \citep{Cha 99} for all the simulations except for the one in Fig.\ref{vrel_rate}, which uses the Hermite code as in previous papers \citep{XZ 08, XZ 09} in order to calculate the relative velocities of planetesimals. As gas drag is dependent on the relative motion between the gas and planetesimals, we must first address the gas motion.

Three possible scenarios can be envisaged for the evolution of a gaseous disk under the perturbations of an inclined companion \citep{Lar 96}. One possibility is that the gaseous disk begins to warp and it is disrupted by differential precession if the disk is extremely thin. Planetesimals would evolve in a gas-free environment which is the case studied by \citet{Mar 09}. Another possibility is that the gaseous disk remains coherent but quickly relaxes to become aligned to the binary orbital plane if the disk viscosity is very high. Then the situation becomes close to that of the coplanar case or slightly inclined case, which have been well studied previously \citep{MS 00, The 06, The 08, The 09, Par 08, XZ 08, XZ 09, Xie 10a, Bea 10, PL 10}. The third possibility which is relevant for the present paper, is that the gaseous disk maintains a coherent structure and precesses as a rigid body ( i.e., $L_{Disk}$ precesses around $L_{Binary}$ as seen in Fig.\ref{orbit}). In addition to the above considerations, the inner structure of the gaseous disk (flow line) would be significantly modified if the binary orbit is eccentric \citep{Kle 00, Par 08, KN 08}.

These effects can result in very complicated gas motions in the disk. In order to understand how gas motion affects the gas drag forces on planetesimals, we first describe four gas disk models (\S\,\ref{gasdisk}), then give the general formula for the gas drag force (\S\,\ref{gasdrag}) and then finally we test the effects of the gas drag resulting from these four gas disk models (\S\,\ref{compare}). Our goal is to obtain a reliable gas drag model for the planetesimals in highly inclined binary systems.


\subsection {Gas Disk}\label{gasdisk}
We consider four models for the gas disk.
\begin{enumerate}
\item The Circular Gas Disk model (CGD), where the gas is in a circular orbit in the circumprimary mid-plane. Following \citet{TL 02}, in cylindrical coordinates (r, z), we adopt a gas-density distribution of the form
\begin{equation}
\rho_g(r, z)=f_d\rho_{g0}\left(\frac{r}{AU}\right)^p {\rm exp}\left(-\frac{z^2}{2h_g^2}\right),
\end{equation}
and a gas velocity
\begin{equation}
V_g(r, z)=V_{k, {\rm mid}}\left[1+\frac{1}{2}\left(\frac{h_g}{r}\right)^2 \left(p+q+\frac{q}{2}\frac{z^2}{h_g^2}\right)\right]
\end{equation}
where $h_g(r) = h_0( r/ {\rm AU} )^{(q+3)/2}$ is the scale height of
gas disk, $V_{k, {\rm mid}}$ is the Keplerian velocity in the
midplane, and $f_d$ is a dimensionless scaling factor. The subscript ``0" indicates the value at 1 AU from the disk center$-$the primary star. In this paper, we take
the Minimum Mass Solar Nebula (hereafter MMSN; \citet{Hay 81}) as
the nominal gas disk, where $f_d=1$, $p = -2.75$, $q = -0.5$, $\rho_{g0} =
1.4 \times 10^{-9}{\rm g. cm^{-3}}$, and $h_0 = 4.7 \times 10^{-2}$
AU.
We note that the CGD model is a rather crude simplification because it completely ignores the effects of the companion star on the gas disk evolution. However, to a first approximation, the CGD is reasonable and it is also the disk model most-often adopted by previous studies \citep{Sch 07, The 06, The 08, The 09, XZ 08, XZ 09}.
\item The Elliptical Gas Disk model (EGD) is the same as the CGD model (same density profile and same flow speed), except that the gas is on an elliptical orbit, with an eccentricity of $e_g$ and precession rate of $\omega_g$. Rigorous values for $e_g$ and
$\omega_g$ require detailed hydro-dynamical modeling on a case-by-case basis, which is beyond the scope of the investigation we wish to perform. Here we follow an alternative strategy and derive simplified analytic expressions for $e_g$ and $\omega_g$, based on the available hydro-dynamical results from previous studies. \citet{Par 08} simulated
$e_g$ and $\omega_g$ for two sets of binaries. Of particular interest are the simulations they performed of a $\gamma$ Cephei-like binary system with mass ratio $q_B=0.234$, orbital semi-major $a_B=20$ AU, and eccentricity $e_B=0.3$. They found that their results for $e_g$ and $\omega_g$ depended on the flux limiter that they used in their code. For the ``minmod'' limiter, the disk was ``quiet" with low eccentricity
$e_g<0.02$ and no precession ($\omega_g=0$). For the ``superbee''
limiter, the disk is ``excited" with $e_g$ up to 0.2 and obvious
precession ($\omega_g \neq 0$). Here we adopt the ``superbee'' results for our
EGD model (the ``minmod" is close to CGD). Using Fig.7 and Fig.8 of \citet{Par 08}, we estimate
\begin{eqnarray}
e_g \sim \left\{ \begin{array}{ll}
0 & \textrm{\ \ if \ \ $r<0.5 $ AU}\\
0.12(r-0.5) & \textrm{\ \ if \ \ $0.5\le r \le 2.5$ AU}\\
0.24-0.02(r-2.5) & \textrm{\ \ if \ \ $r > 2.5$ AU},
\end{array} \right.
\label{eg}
\end{eqnarray}
and
\begin{equation}
\omega_g \sim 2\pi/1500 {\rm yr}.
\label{wg}
\end{equation}
(personal communication with Paardekooper reveals that the original data underlying these simulations has been lost due to disk death, hence the need for us to create an approximate reconstruction directly from the published paper).

We note that, in fact, the disk will never reach a steady state (see the Fig.8 of Paardekooper et al. 2008), and thus the above estimation of $e_g$ and $\omega_g$ should be considered as an averaged approximation. Note also that Eqn.\ref{eg} and Eqn.\ref{wg} are valid only for the $\gamma$ Cephei-like case, a more generic expression for $e_g$ and $\omega_g$ which includes the dependency on the binary configuration would require more simulation results based on hydro-dynamical modeling.

\item  The Circular Gas Disk + Nodal Precession model (CGD+N), has the same inner structure as the CGD model, but adds in the precession of the ascending
node of the disk plane relative to the binary orbital plane.
Nodal precession occurs when the disk plane has a
considerable tilted angle ($i_B$) relative to the binary orbital
plane. \citet{Lar 96} found that, if the Mach number of the
disk is not too large ($Ma < 30$ roughly), then the disk can
maintains its structure and undergoes a near rigid body precession at a
rate given by
\begin{equation}
\Omega_g=-\frac{15M_B R_d^3}{32M_A a_B^3}{\rm cos}i_B\Omega_k,
\end{equation}
where $M_A$ and $M_B$ are the masses of the primary and secondary
stars, $\Omega_k$ is the local Keplerian rotation rate and $R_d$ is
the radial size of the truncated disk. For the $\gamma$ Cephei like
case, we set $R_d=4$ AU according to \citet{AL 94} and \citet{Pic 05}.

\item The Eccentric Gas Disk + Nodal Precession model (EGD+N), has the same
inner structure as the EGD model and the same nodal precession as the CGD+N model.
\end{enumerate}

From physical considerations, EGD+N is the most realistic disk model of the four.
However, as we will see in \S\,\ref{compare}, the four disk models all
result in similar gas drag effects in highly inclined cases.

\subsection{Gas Drag}\label{gasdrag}

A spherical particle of radius $R_p$, moving through a gas with a relative velocity $V_{rel}$, experiences an aerodynamic drag force $F_d$ given by \citep[e.g.][]{Arm 10}
\begin{equation}
{\bf F_d}=-\frac{C_D}{2}\pi \rho_g R_p^2 V_{rel} {\bf V_{rel}},
\label{fd}
\end{equation}
where $\rho_g$ is the gas density and $C_D$ is the drag coefficient.
The form of the drag coefficient depends upon the size of the particle compared to the mean free path
$\lambda$ of molecules in the gas ($\lambda\sim 0.27 (r/AU)^{-11/4}$
cm for the MMSN). If $R_p<9/4\lambda$, then
\begin{equation}
C_D=\frac{8}{3}\left(\frac{8}{\pi}\right)^{1/2}\frac{C_S}{V_{rel}},
\end{equation}
where $C_S\sim\Omega_k h_g$ is the local sound speed. This is called
the Epstein regime of drag. For larger particles the Stokes drag law
is valid, and the drag coefficient can be expressed as a piecewise
function
\begin{eqnarray}
C_D = \left\{ \begin{array}{ll}
24Re^{-1} & \textrm{\ \ if \ \ $Re<1$}\\
24Re^{-0.6} & \textrm{\ \ if \ \ $1\le Re\le800$}\\
0.44 & \textrm{\ \ if \ \ $Re>800$.}
\end{array} \right.
\end{eqnarray}
Here $Re=2R_p\rho_gV_{rel}/ \mu$ is the Reynolds number, and
$\mu=(8/\pi)^{1/2}\rho_gC_s\lambda/3$ is the kinematic viscosity of
the gas.
%
%
%
%

\subsection{Model Test and Comparison}\label{compare}
All the tests and comparisons are conducted in the $\gamma$ Cephei-like case as in \citet{Par 08} with mass ratio
$q_B=0.234$, orbital semimajor $a_B=20$ AU, and eccentricity
$e_B=0.3$. Although such a binary configuration is too tight to be relevant for the highly inclined case (usually $a_B>$30-40 AU according to \citet{Hal 94}), we still choose it as our test case because hydrodynamic results are currently available only for such close binaries \citep{Par 08}.
In the following, we consider two sets of tests or comparisons.

\begin{figure}
\begin{center}
\includegraphics[width=\textwidth,bb=0 0 800 800]{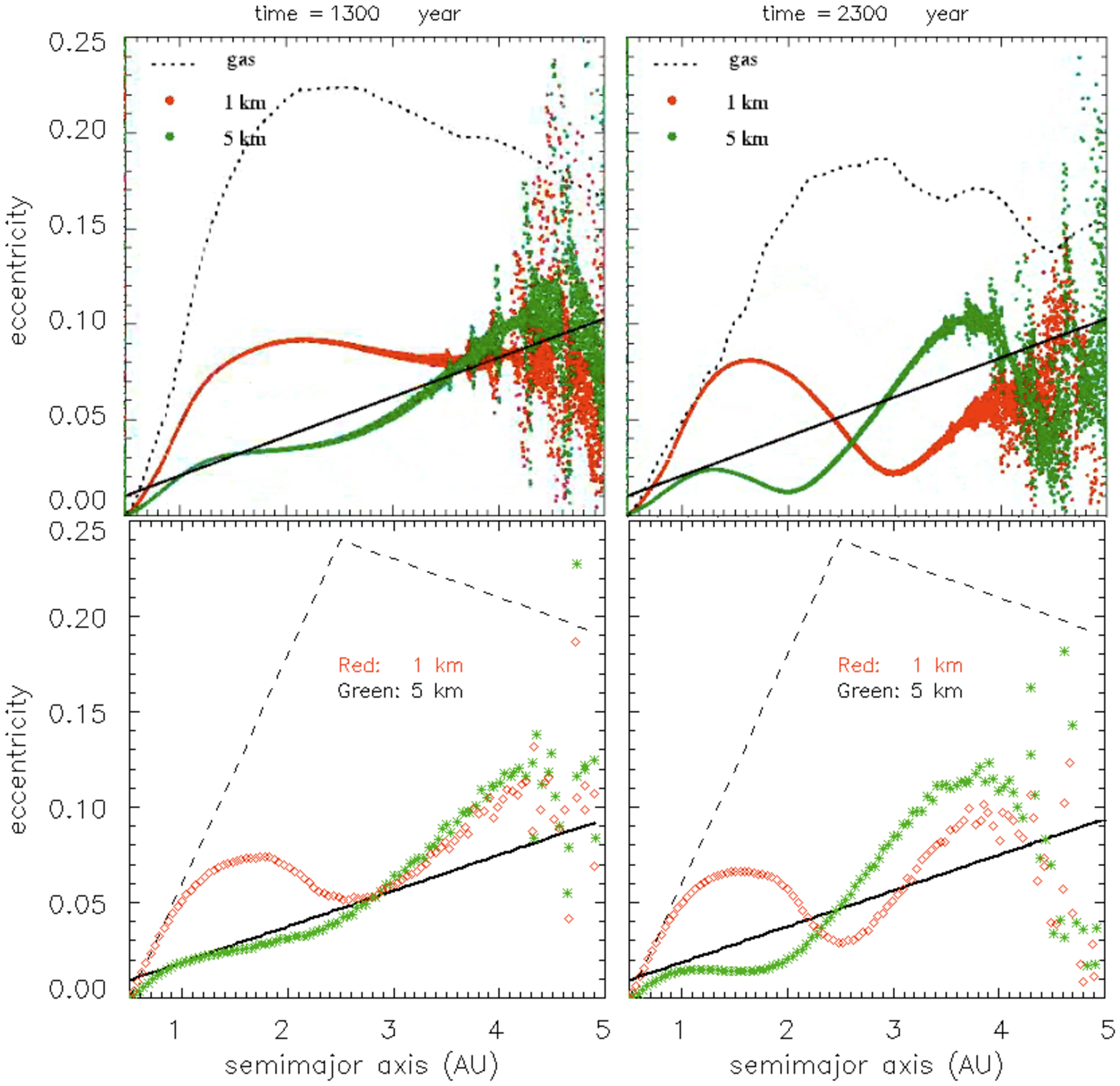}
  \caption{Comparison between the results given by the hydrodynamical model (top: \citet{Par 08}) and the approximate EGD model (bottom). We display snapshots of planetesimal eccentricity vs. semimajor at t=1300 and 2300 yr. The black solid lines indicate the forced eccentricity $e_f=(5ae_B)/(4a_B)$. The black dashed lines show the eccentricity distribution of the gas. We see that the results from the EGD model can recover the basic features of the results by \citet{Par 08}, thus providing us with a means to simply and quickly add the main features of the time-consuming hydrodynamical results into a fast N-body code.
  }
  \label{com_par08}
   \end{center}
\end{figure}

\begin{figure}
\begin{center}
\includegraphics[width=\textwidth]{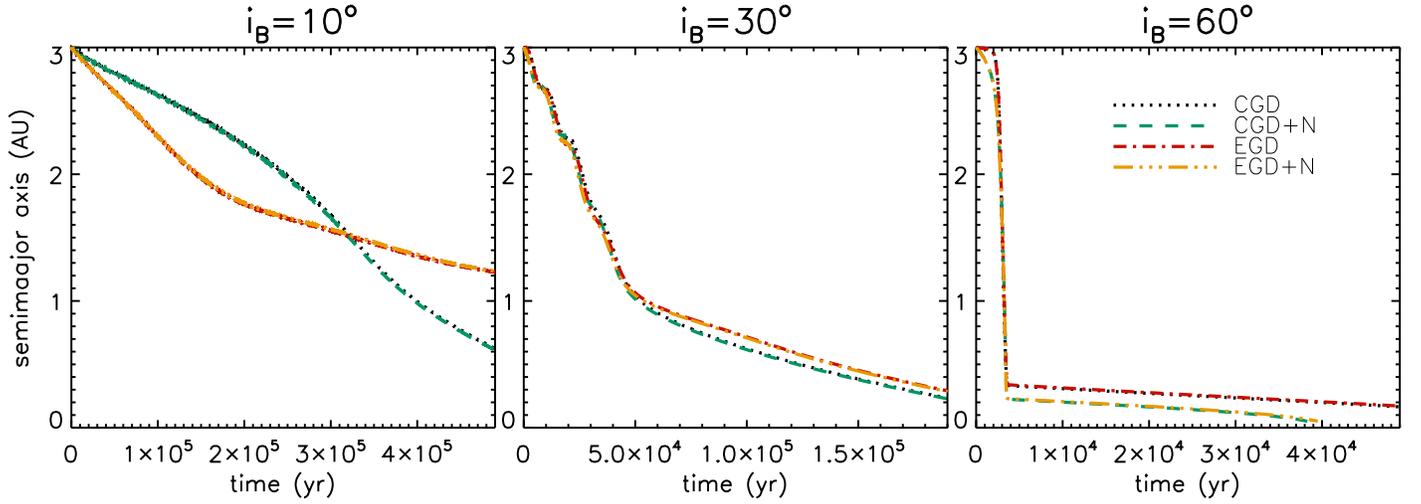}
  \caption{Comparison between the results obtained from 4 different disk models (CGD, CGD+N, EGD, EGD+N) for three cases with $i_B=10^\circ, 30^\circ$ and $60^\circ$. We plot the evolution of a planetesimal's semi-major axis. We see, for highly inclined cases ($i_B>30^\circ$), the evolution of planetesimal semi-major axis is nearly independent of the eccentricity and nodal precession of the gas. We thus feel justified in using the simple CGD model for the remainder of our investigation due to the fact that we are concentrating on the high inclination cases.
 A detailed investigation of the planetesimal dynamics, including the reason for large inward jumps is provided in \S\,\ref{SECN:SINGLE}.
}
  \label{com_4}
   \end{center}
\end{figure}

\begin{itemize}
\item We first test the EGD model through comparison (see Fig.\ref{com_par08}) to the hydro-dynamical simulations of \citet{Par 08}.  In our simulation with the EGD model, 100 particles with $R_p=1$ and $R_p=5$ km respectively were initially set in circular
orbits in the coplanar plane with semi-major axes from 0.5 to 5 AU.
As can be seen in Fig.\ref{com_par08}, the EGD model can roughly recover the basis features of Paardekooper's results, such as (1) the shape of the eccentricity profile of planetesimals, and (2) the trend that smaller particles ($R_p=1$ km) behavior more closely matches that of the gas in the inner regions of the disk (because gas drag is more efficient for smaller size particles and denser gas density). Therefore, we believe that the EGD model, though rather simplified, provides a reasonable means to include the basic effects of gas-disk eccentricity induced by the companion's perturbations.

\item We then compare the results of the four gas disk models (see Fig.\ref{com_4}). We set three planetesimals with radial size $R_p=5$ km, and with circular
orbits starting at 3 AU (i.e., the initial semimajor axis $a_{p0}$=3 AU) in the gas disk plane. We consider a near-coplanar case ($i_B=10^\circ$), an intermediately inclined case ($i_B=30^\circ$) and a highly inclined case ($i_B=60^\circ$)
for which the Kozai effect should be apparent. As can be seen in Fig.\ref{com_4}, only for the near-coplanar case does the gas eccentricity have any
  significant effect on the evolution of the planetesimal semi-major damping. That is because the gas drag force depends mainly on the planetesimal's orbital inclination with respect to the gas disk plane in significantly inclined binary cases (see also in Fig.\ref{aei1} and Fig.\ref{aei2}).  In addition, we see that the nodal precession of the gas disk has little effect for all the cases. This is well understood as the planetesimal node always undergoes nodal precession independent of  whether the nodal precession of the gas disk occurs or not.
A detailed investigation of the planetesimal dynamics, including the reason for large inward jumps evident in Fig.\ref{com_4} is provided in \S\,\ref{SECN:SINGLE}.

\end{itemize}

In summary, through the above comparisons, we see that the most
commonly adopted model, i.e., the CGD model, is only a rather crude
simplification in the near coplanar cases ($i_B<10^o$), but it is a
good approximation in highly inclined cases ($i_B\ge30$). This conclusion, although obtained from a test on a specific binary with
a $\gamma$ Cephei-like configuration, should apply
 to binaries with any separation since there is not any significant separation dependence as seen from the above test and comparison. The EGD+N model, though more realistic, is currently available only for close binaries with separation of 10-20 AU, while the CGD model can readily be applied to any case. Therefore, given our interests in this paper in modeling highly inclined binaries with separation of  $>$30-40 AU, hereafter we adopt the CGD as our standard gas-disk model, and focus on it as we try to understand the planetesimal behavior in binary systems with different configurations ($a_B, e_B, i_B$) and mass ratios ($q_B$).

\section{Dynamics of Individual Planetesimals: Inward Jumping}\label{SECN:SINGLE}
\subsection{Different Regimes of Semimajor Axis Damping}\label{damp_regime}
Before presenting detailed numerical results, we present in this chapter analytical considerations in order to  understand the different regimes in which gas drag can operate on planetesimals in a circumprimary disk.

The main effect of gas friction is to force an inward drift of planetesimals.
According to \citet{ada 76}, the average migration rate (or the
semi-major axis damping rate) follows the approximate relation
\begin{eqnarray}
\frac{\tau_0}{a}\widehat{\frac{da}{dt}}&\sim &-2\left(\frac{5}{8}\hat{e}^2+\frac{1}{2}\hat{i}^2+\eta^2\right)^{1/2} \nonumber \\
& &\times\left(\eta+\frac{17}{16}\hat{e}^2+\frac{1}{8}\hat{i}^2\right),
\label{adamp}
\end{eqnarray}
where $a$, $e$, $i$ are the orbital semimajor axis, eccentricity and
inclination (\emph{relative to the disk plane rather than the binary orbit plane}) of the planetesimal, the hat symbol (e.g. $\hat{e}$ and $\hat{i}$) denotes the average value of that variable taken over at least one orbital period,  $\eta$ denotes the deviation of the gas rotation from the Keplerian velocity, and $\tau_0$ is a characteristic timescale
depending on the gas density and planetesimal radial size. For
a gas disk scaled in proportion to the MMSN,
we have $\eta\sim2\times10^{-3}$ and $\tau_0$ can be written as
\begin{equation}
\tau_0 \sim 24f_d\left(\frac{\hat{i}}{h_0}\right)\left(\frac{R_p}{\rm km}\right)\left(\frac{a}{\rm AU}\right)^{3}\left(\frac{M_A}{M_{\odot}}\right)^{-1/2} \ \  \rm yr,
\label{tau}
\end{equation}
where $f_d$ is the scaling number of the disk density compared to
MMSN, and the factor $(\hat{i}/h_0$) accounts for the decrease in the gas density in the vertical direction.
%

Eq. \ref{adamp} indicates that the migration rate has contributions from
three sources. One is $\eta$, which accounts for the fact that the pressure supported gas disk departs from purely Keplerian dynamics and only depends on the properties of the
gas disk. The other two are the planetesimal orbital eccentricity
($\hat{e}$) and inclination ($\hat{i}$), which can be excited via various mechanisms by the companion star, and measure the angle between planetesimal orbits and gas streamlines.  If, for example, $\hat{i}\gg\hat{e}\, {\rm and} \, \eta$, then Eq.\ref{adamp} gives $da/dt\propto \hat{i}^2$; planetesimals undergo a much faster inward migration with increasing $\hat{i}$. This first order estimate indicates that
highly inclined planetesimals never evolve in a manner close to the gas-free case, despite the fact that they spend more than $80-90\%$ of their orbital time in an essentially gas-free environment \citep{Mar 09}.

\begin{figure}
\begin{center}
\includegraphics[width=\textwidth]{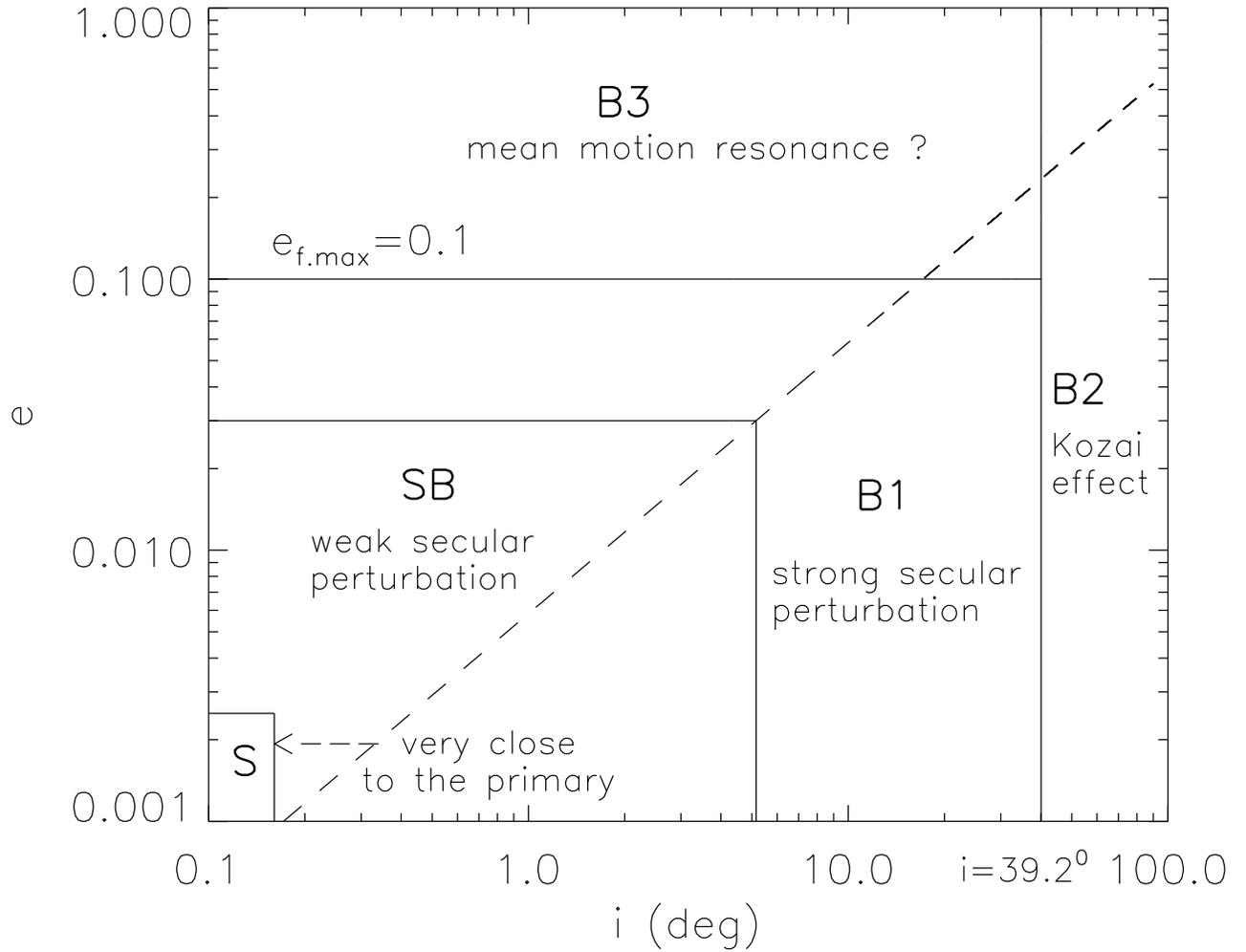}
  \caption{Schematic diagram of different regimes for the damping of
planetesimal semimajor axis. The damping is dominated by
$\eta$ (the sub-Keplerian gas velocity) in region S, the evolution of $i$ and $e$ dominates the gas damping in region B. Region B is divided into 3 sub-regions depending on the particular mechanism by which $e$ and $i$ are excited; secular perturbations (B1),
Kozai effect (B2), and other effects such as mean motion resonance by the companion star (B3). SB is a transition region between S and B, in which contributions from $e$, $i$ and $\eta$ can all be significant. The dashed line divides the plot into two parts: $e$ has more contribution to the damping of semi-major axis than than $i$ does in the top-left part, while in the bottom-right part the opposite is true. For full details, see the text in \S\,\ref{damp_regime}}
    \label{ei}
   \end{center}
\end{figure}

Fig.\ref{ei} is a schematic diagram showing which source or mechanism dominates the inward migration. Three main regions, corresponding to different regimes for gas drag, can basically be distinguished:
(1) region S, where planetesimal migration is close to that in an unperturbed, single star case,
(2) region B where planetesimal migration is dominated by the binary perturbations, and
(3) region SB, a transition region between B and S.
In addition, region B can be divided into three subregions corresponding to three different mechanisms which can excite the $\hat{e}$ and/or $\hat{i}$.
The precise definitions of all these regimes is as follows:
 \begin{itemize}
 \item Region S: where $\frac{5}{8}\hat{e}^2<\eta^2 \, {\rm and} \,
 \frac{1}{2}\hat{i}^2<\eta^2$. For this low dynamical excitation case, according to Eq.\ref{adamp}, the migration rate is dominated by $\eta$. For MMSN-like gas disks, Region S corresponds to planetesimals that are very close to the primary as compared to the distance of the companion, i.e. $a/a_B\ll1$. In this case, $\hat{e}$ and $\hat{i}$ remain very small and planetesimal orbits are close to what they would be in an unperturbed single-star case. Note here that the inward migration rate (Eqn.\ref{adamp}) $\widehat{da/dt}\propto \eta^2a/\tau_0\propto a^{-1}$, meaning that the inward migration is accelerated in region S.

 \item Region B: $\hat{e}$ and/or $\hat{i}$ are significantly pumped up by the companion's perturbations and thus dominate the damping rate of planetesimal semimajor axis. Depending on the specific mechanism that pumps up $\hat{e}$ and/or $\hat{i}$, region B is divided into three subregions:
 \begin{itemize}
 \item subregion B1: where $\hat{e}$ and/or $\hat{i}$ are excited by secular
perturbations, but Kozai effects are absent, i.e., $i_B<39.2^\circ$. If we assume the average values of $\hat{e}$ and $\hat{i}$ approximate their forced values $e_f$ ($e_f<\sim0.1$, see the Appendix) and $i_f$ ($i_f\sim i_B<39.2^\circ$), then region B1 corresponds to the middle part of Fig.\ref{ei}.
 \item subregion B2, the Kozai regime, i.e., $\hat{i}\sim i_B>39.2^\circ$. In such case, both $\hat{e}$ and $\hat{i}$ can be highly excited via the Kozai effect.
 \item subregion B3, where $\hat{e}$ is excited and it is independent of $\hat{i}$.
This region may correspond to a mean motion resonance scenario. An example can be found in \citet{IL 96}, which shows the combination of gas drag and Jupiter's mean motion resonances can affect the structure of the asteroid belt.
\end{itemize}

 \item Intermediate region SB: where  $\frac{5}{8}\hat{e}^2>\eta^2 \, {\rm and} \, \frac{1}{2}\hat{i}^2>\eta^2$ but $\frac{17}{16}\hat{e}^2<\eta \, {\rm and} \,
\frac{1}{8}\hat{i}^2<\eta$. Region SB corresponds to the transition
case, where $\hat{e}$ and $\hat{i}$ are only mildly excited, meaning that the
migration rate is determined by the combination of $\eta$,
$\hat{e}$ and/or $\hat{i}$.
\end{itemize}

The dashed line in Fig.\ref{ei}, i.e., $\frac{17}{16}\hat{e}^2=\frac{1}{8}\hat{i}^2$,
divides the entire plot into two parts: In the upper-left $\hat{e}$ dominates the damping of semi-major axis, while in the bottom-right of the plot $\hat{i}$ contributes more than $\hat{e}$. We define the boundary between regions B1 and B3 as being $\hat{e}=0.1$, since the maximum forced eccentricity ($e_f$) for a typical binary system is $\sim 0.1$ (see the Appendix for detail).

Note that Fig.\ref{ei} is based on Eq.\ref{adamp} which is derived
from the most simplified gas drag model, i.e. the circular gas disk
model (CGD). According to the test in \S\,2.3 (see also in
Fig.\ref{com_4}), the CGD model is a good
approximation only for relative large $\hat{i}$. Therefore, hereafter we
focus only on these large $\hat{i}$ (or $i_B$) cases, and simply classify them into two categories: Kozai-off (which corresponds to the right-hand section of the B1 region) and Kozai-on (the whole of the B2 region). We don't explore region B3 in the present paper, as it is likely only to be relevant to some isolated (i.e., in resonance) or unstable regions (i.e., close to the companion star) of the protoplanetary disk.

\subsection{Kozai-Off Regime}\label{off-kozai}
In the following sections, we turn to numerical simulations using the circular gas disk model (CGD) which was tested and justified in \S\,\ref{SECN:MODEL}.

We first define our standard system parameters upon which we base most of our variations. We take $M_A=2M_B=M_\odot$, $a_B=50$ AU, $e_B=0.3$, $i_B=30^o$, $f_d=1.0$, and $R_p=5$ km. The mass ratio and eccentricity are chosen as the typical values for known binaries \citep{DM 91}. The separation is chosen as it is wide enough to be relevant for large inclination and tight enough to have the binary companion cause considerable perturbations in the $0.1 - 10$ AU region, where the core-accretion model of planet formaiton prefers. $i_B=30^o$ is chosen as it is large enough for the validity of CGD model as tested in \S\,\ref{compare}, but not so large as to induce the Kozai effect. Such a binary configuration leads to a stable orbit limit of $\sim10$ AU around the primary star \citep{HW 99}.

\begin{figure}
\begin{center}
\includegraphics[width=0.65\textwidth]{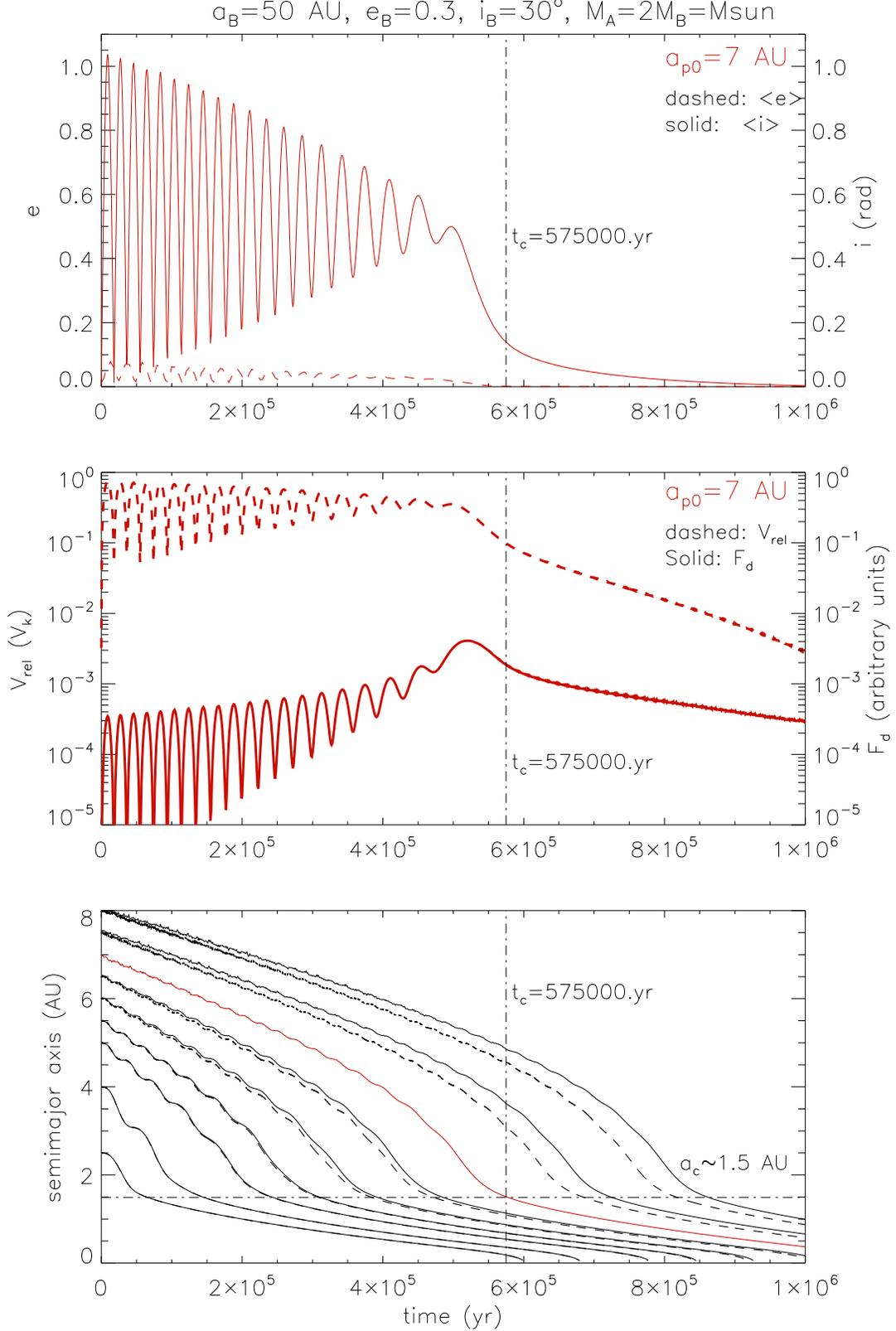}
 \caption{
 The evolution of orbital eccentricity ($\hat{e}$, dashed, top), inclination ($\hat{i}$, solid, top), velocity relative to the gas ($V_{rel}$ in units of local Keplerian velocity $V_k$ , dashed, middle), gas drag force ($F_d$, solid middle), semimajor axis (red solid, bottom) of a planetesimal with $a_{p0}=7$ AU, and $R_p=5$ km under the coupled effects of perturbations from a binary ($a_B=50$ AU, $e_B=0.3$, $i_B=30^\circ$, $M_A=2M_B=M_\odot$) and gas drag from a gas disk (1 MMSN) . Planetesimal orbital eccentricities and inclinations are initially set as very small values ($<10^{-4}$), and their other initial angular orbital elements are set as random values from $0$ to $360^\circ$.
In the bottom panel, we also plot the results for planetesimals with
different initial semi-major axis $a_{p0}$. The dashed curves are
the results of retrograde case ($i_B=150^\circ$) for comparison. We
see that all planetesimals' inward migrations transit from a fast
regime to a slow one at a similar turn-over semi-major axis, which
is independent of $a_{p0}$. This turn-over semi-major axis can be
well described with $a_c$ (Eq.\ref{ac} and see also the horizonal
dot-dash line in the bottom panel).
Note that $\hat{i}$ is measured relative to the plane of the gas disk rather than to the binary orbital plane.
%
The vertical dot-dash line denotes the time at the turnover point
for the case of $a_{p0}=7$ AU.
%
}
      \label{aei1}
   \end{center}
\end{figure}

This high-$\hat{i}$ case corresponds to the right-hand side of the B1 region in
Fig.\ref{ei}. Fig.\ref{aei1} presents the result of a numerical integration showing the typical behavior of a planetesimal orbit in this regime (see also in the middle panel of Fig.\ref{com_4}- but note different binary parameters).
As expected from linear secular perturbation theory, $\hat{e}$ and
$\hat{i}$ undergo oscillations with a period equal to the secular perturbation
time scale, i.e. $t_{sp}$.  The value of $t_{sp}$ derived from Fig.\ref{aei1} is in good agreement with the  first order approximation derived by \citet{The 06}:
\begin{eqnarray}
t_{sp}\sim \frac{4}{3}\left(\frac{a}{\rm AU}\right)^{-3/2}\left(\frac{a_B}{\rm AU}\right)^3(1-e_B^2)^{3/2} \nonumber \\
\left(\frac{M_A}{M_{\odot}}\right)^{1/2}\left(\frac{M_B}{M_{\odot}}\right)^{-1} \ \ \ {\rm yr},
\label{tsp}
\end{eqnarray}
where $M_A$, $M_B$, and $M_{\odot}$ are the masses of the primary,
secondary and the Sun respectively.
As the relative velocity $V_{rel}$ between the planetesimal and gas can be approximately estimated as (see Eqn 4.20 of \citet{ada 76})
\begin{eqnarray}
V_{rel}^2\sim V_k^2(\frac{5}{8}\hat{e}^2+\frac{1}{2}\hat{i}^2+\eta^2),
\label{vrel}
\end{eqnarray}
this means that in the regime where $\hat{i}\gg \hat{e} ({\rm and}\, \eta)$, $V_{rel}$ and the gas drag force $F_d$ (Eqn.\ref{fd}) will also follow the same periodic oscillation as $\hat{i}$. This oscillation of gas drag damping causes the planetesimal semimajor axis to damp step-by-step as seen in the bottom panel of Fig.\ref{aei1}.

In addition, as can be seen in Fig.\ref{aei1}, the
oscillation amplitudes of $\hat{i}$ and $\hat{e}$ are under-damping because of gas drag. According to
\citet{ada 76}, the damping of $\hat{i}$ and $\hat{e}$ follows,
\begin{eqnarray}
\frac{\tau_0}{\hat{e}}\widehat{\frac{de}{dt}} & \sim & 2\frac{\tau_0}{\hat{i}}\widehat{\frac{di}{dt}}\sim -\left(\frac{5}{8}\hat{e}^2+\frac{1}{2}\hat{i}^2+\eta^2\right)^{1/2}.
\label{eidamp}
\end{eqnarray}
In the present case, where $\hat{i}\gg \hat{e}\,({\rm and}\, \eta)$, the timescale of $\hat{i}$ damping can thus be approximated by
\begin{eqnarray}
t_{dp}\sim\frac{\hat{i}}{\widehat{di/dt}}\sim2^{3/2}\frac{\tau_0}{\hat{i}}.
\label{tdp1}
\end{eqnarray}
Substituting Eq.\ref{tau} into Eq.\ref{tdp1}, $t_{dp}$ can then be rewritten as
\begin{eqnarray}
t_{dp}\sim1444f_d^{-1}\left(\frac{R_p}{\rm km}\right)\left(\frac{a}{\rm AU}\right)^3\left(\frac{M_A}{\rm M{\odot}}\right)^{-1/2} \ \ \ {\rm yr}
\label{tdp2}
\end{eqnarray}

Depending on whether the secular perturbation or the gas drag damping dominates, Fig.\ref{aei1} can be divided into two parts with the boundary between these two regions being roughly set by the last turn-over point of the migration curves: E.g. for the red-curve in Fig.\ref{aei1}, the coordinates of the critical time and semi-major axis at which this occurs is  $t_c\sim5.75\times10^5$ yr and $a_c\sim1.5$ AU for $a_{p0}= 7$ AU.
For times earlier than $t_c$ or semi-major axes beyond $a_c$, $\hat{i}$ is periodically excited, and thus the planetesimal periodically undergoes a step-wise inward migration with the steep drops in semi-major axis corresponding to the peaks in $\hat{i}$. For times greater than $t_c$ or semi-major axes inside $a_c$, $\hat{i}$ is smoothly damped to a low value, thus the planetesimal migrates inward smoothly and slowly.

We can also estimate $a_c$ by equating $t_{sp}$ and $t_{dp}$ (Eq.\ref{tsp} and Eq.\ref{tdp2}),
which gives a ``zeroth-order" estimation of $a_{c0}$ as,
\begin{eqnarray}
a_{c0}\sim0.9\times f_d^{2/9}\left(\frac{a_B}{\rm 10 AU}\right)^{2/3}(1-e_B^2)^{1/3} \nonumber \\
\left(\frac{R_p}{\rm km}\right)^{-2/9}\left(\frac{M_A}{M_B}\right)^{2/9} \ \ {\rm AU}.
\label{ac0}
\end{eqnarray}
We call $a_{c0}$ ``zeroth-order" because this timescale analysis gives the correct dependency of $a_c$ on the various parameters (for example, $a_c$ is independent of $a_{p0}$ in Fig.\ref{aei1}), but not the correct absolute coefficient. Through many sets of simulations, such as Fig.\ref{aei1} but with different stellar mass ratios ($M_A/M_B=$0.2, 0.5, 1.0), binary semimajor axies ($a_B=$10, 20, 50, 100, 200 AU), binary eccentricities ($e_B=$0, 0.3, 0.6), binary inclinations ($i_B=20^\circ, 30^\circ, 39^\circ$ and $150^\circ$), gas disk densites ($f_d=$0.05, 0.5, 1), and planetesimal radial sizes ($R_p=$1, 5, 100 km), we found similar results for both the prograde and retrograde cases (see the bottom panel of Fig.\ref{aei1}) and the empirically measured form of $a_c$ is just systematically smaller by $\sim 33\%$ than $a_{c0}$, i.e.,
\begin{eqnarray}
a_{c}&\sim&\frac{2}{3}a_{c0}\nonumber \\
&\sim& 0.6\times f_d^{2/9}\left(\frac{a_B}{\rm 10 AU}\right)^{2/3}(1-e_B^2)^{1/3} \nonumber \\
&&\left(\frac{R_p}{\rm km}\right)^{-2/9}\left(\frac{M_A}{M_B}\right)^{2/9} \ \ {\rm AU}.
\label{ac}
\end{eqnarray}
We note that $t_{dp}$ is larger by just about one order of magnitude than $t_{sp}$ at $a_c$.

The physical meaning of $a_c$ is that it indicates the turn-over point where the planetesimal semi-major axis transits from the excited (usually fast) damping regime to the quiet (usually slow) damping regime (also corresponding to the boundary between the region B1 and SB). The importance of $a_{c}$ lies in the fact that it determines the location in the proto-planetary disk where planetesimals will pile up (see \S\,\ref{SECN:SWARM} for details).


\subsection{Kozai-On regime}\label{on-kozai}
The standard setup for this case (corresponding to the B2 region in Fig.\ref{ei}) is chosen as similar to that in \S\,\ref{off-kozai}, but with $i_B=60^\circ$ instead.  As $i_B$ is larger than the critical value $39.2^\circ$©V(R), the Kozai effect is turned on (Kozai 1962), and $\hat{e}$ can be excited to a degree which is comparable with $\hat{i}$. In such a case, both $\hat{e}$ and $\hat{i}$ have major contributions to the planetesimal migration rate (see Eqn.\ref{adamp}).

The most interesting result is that large eccentricity oscillations can lead to very fast inward migration of the planetesimals towards the inner disk. As can be seen
in Fig.\ref{aei2}, a planetesimal starting at $a_{p0}=3$ AU spirals down to 1\,AU in $\sim 10^{4}\,$years. For the $a_{p0}=5$ AU case, the planetesimal starts closer to the companion star and is subjected to much stronger Kozai perturbations. Eccentricities can quickly reach values as high as 0.7 that places the planetesimal pericenter
in inner gas rich regions where it will in addition reach very high $V_{rel}$ values. This leads to a very fast, almost ``jump"-wise inward migration from 5 AU to $\sim0.35$ AU.
For the $a_{p0}=8$ AU case, the eccentricity also reaches 0.7 but, due to the larger initial semi-major axis, the planetesimal cannot directly reach very dense inner regions and thus first undergoes several small jumps, until finally making a bigger jump to $\sim 1$ AU.

All the inward jumps occur when $\hat{e}$ and/or $\hat{i}$ is excited,
leading to a surge in the experienced gas drag force ($F_d$). Although the $\hat{e}$ contribution to $F_d$ is significant, it seems that $\hat{i}$ is still the major component of $F_d$. Note $\hat{i}$ is with respect to the disk plane rather than to the binary orbital plane, and thus the periodic variation of $\hat{i}$ shown in Fig.\ref{aei2} is caused by the Nodal precession of the planetesimal orbit with a period of two times that of the Kozai cycle period.

For all cases, once the planetesimals complete their jump, they will approximately stabilize at that semi-major axis, with very small residual $\hat{e}$ and $\hat{i}$ and thus only very slow inward migration. They no longer suffer from the Kozai effects because they have jumped into an inner region where the gas is so dense that gas drag damping dominates over Kozai excitation. Unlike the Kozai-off case, the turn-over semi-major axis, where the planetesimal stops jumping, is no longer independent of $a_{p0}$ and not equal to $a_c$ (Eqn.\ref{ac}). Fig.\ref{aei2_turn} shows how it varies with $a_{p0}$ in the
Kozai-on case.
\begin{itemize}
\item If $a_{p0}\lesssim1.5$ AU, the turnover semi-major axis is around  $a_{p0}$, which means that no ``inward-jumping" occurs to the planetesimals in the very inner region; Kozai effects is fully turned off because such dense gas and hence strong gas drag damping dominates over Kozai excitation.
\item If $1.5\lesssim a_{p0}\lesssim 8$ AU, the turn-over semi-major axis follows a characteristic oscillatory pattern, whose amplitude varies between a lower limit at $a_{c(low)}\sim0.3$ AU and an upper limit at $a_{c}\sim 1.5$ AU roughly;
\item If $a_{p0}\gtrsim 8$ AU, the planetesimal is initially located close to the boundary of the stable orbit ($\sim 10$ AU), thus the results become chaotic. However most, if not all, of the turn-over points remain confined within the range between $a_{c(low)}$ and $a_c$, i.e., 0.3-1.5 AU.
\end{itemize}

In Fig.\ref{fac}, we study the dependence of the lower limit of  the turn-over semi-major axis ($a_{c(low)}$) on $a_B$ and $i_B$. Empirically we found that
\begin{eqnarray}
a_{c\,(low)} \sim \left\{ \begin{array}{ll}
0.50a_{c} & \textrm{\ \ if \ \ $i_B=50^{\circ}$}\\
0.20a_{c} & \textrm{\ \ if \ \  $i_B=60^{\circ}$}\\
0.10a_{c} & \textrm{\ \ if \ \  $i_B=70^{\circ}$}
\end{array} \right.
\label{ac_low}
\end{eqnarray}
 where $a_{c}$ is given by Eqn.\ref{ac}. For even larger $i_B$($70^\circ -90^\circ$), the $a_{c(low)}$ turns to be larger. The $a_{c(low)}$ for prograde and retrograde cases are symmetrical about $i_B=90^\circ$.
 The combination of $a_{c(low)}$ and $a_c$ determines a region for the
Kozai-On case, into which the planetesimals are likely to jump and pile up.

\section{Behavior of a swarm of planetesimals: Piling Up}\label{SECN:SWARM}

\subsection{Initial Set-up}\label{setup}

In \S\,\ref{SECN:SINGLE}, we focused solely on the dynamics of individual planetesimals. In this section, we investigate the behavior of a swarm of planetesimals, focusing on the evolution of the planetesimal disk's surface density $\Sigma$, under the coupled effects of gas drag and binary perturbations.

In each simulation, all planetesimals are initially set onto a near-coplanar circlular orbit around the primary star with orbital eccentricity and inclination (relative to the gas disk plane) $0\le e=2i\le10^{-4}$. All other angular elements are randomly distributed. The planetesimals' semimajor axes are uniformly distributed in a range ,i.e., $a_{in}<a<a_{out}$ with the same radial size ($R_p$). We set $a_{in}=0.05$ AU, within which planetesimals are removed from the simulation.  $a_{out}$ is set as the boundary of the stable orbit according to \citet{HW 99}.  In order to obtain reliable statistical results, we initially put $N_{p0}=100$ planetesimals in every semimajor axis bin of 1 AU width. By monitoring the planetesimal number ($N_p$) in each bin, we can estimate the change of planetesimal surface density, i.e., $\Sigma /\Sigma_0$.

Note that during the statistical process (i.e., counting $N_p$), we asign two weighting factors ($w_1$, $w_2$) to each planetesimal (the tracked test particles), where $w_1=a_{p0}^{-1/2}$ leading to an initial surface density profile of $\Sigma_0\propto a_{p0}^{-3/2}$ as in MMSN, and $w_2=4.2$ if $a_{p0}>2.7$ AU otherwise $w_2=1$, accounting for the solid density enhancement beyond the snow line.
The gas disk model is the CGD model as described in \S\,\ref{gasdisk}.

We conduct a large set of simulations in which we vary the stellar mass ratios ($q=M_B/M_A=0.2, 0.5, 1.0$), binary orbital semi-major axes ($a_B=10, 20, 50, 100, 200$ AU), binary orbital eccentricities ($e_B=0, 0.3, 0.6$), binary orbital inclinations ($i_B=10^\circ,30^\circ, 50^\circ, 60^\circ, 70^\circ, 80^\circ, 90^\circ, 100^\circ, 110^\circ, 120^\circ, 130^\circ, 150^\circ$), and the planetesimal radial sizes ($R_p=5, 100$ km).  For the convenience of describing the results, we first define a standard case, then we vary one of parameters of the standard case each time to see the effects of the parameter on the result.

\subsection{the Standard Case}\label{standard}
\begin{figure}
\begin{center}
\includegraphics[width=0.8\textwidth]{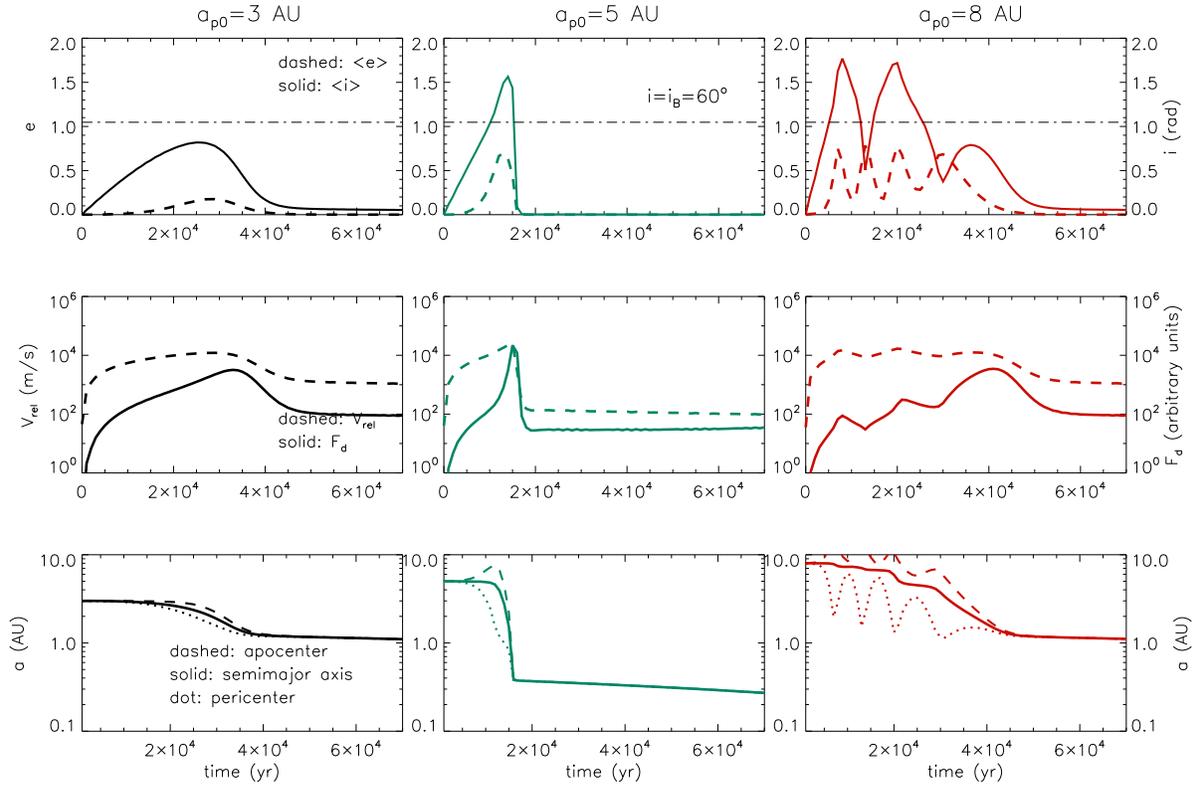}
  \caption{The same as Fig.\ref{aei1}, but for the Kozai-on case with $i_B=60^\circ$.  We see that high $\hat{i}$ and $\hat{e}$ (induced by the Kozai
effect) significantly increase the inward migration; planetesimals
can even ``jump" into the very inner region, and unlike to that in the Kozai-off case, we now find that the turn-over semi-major axis (where the jumping stops) is \emph{no longer} independent on $a_{p0}$ (see Fig.\ref{aei2_turn}).
%
The horizonal dot-dash line denote $i=i_B=60^\circ$.
 }
    \label{aei2}
   \end{center}
\end{figure}

\begin{figure}
\begin{center}
\includegraphics[width=\textwidth]{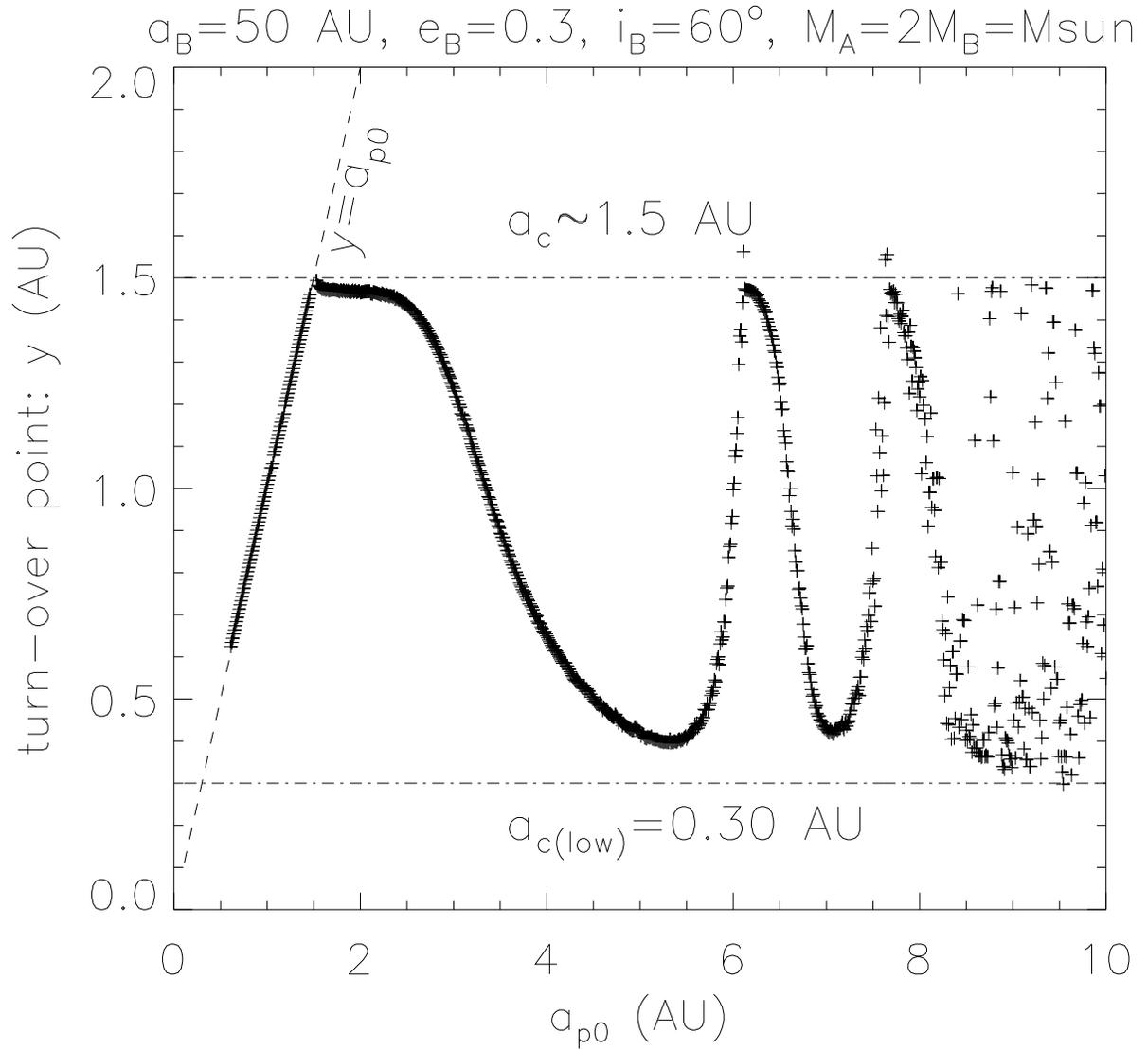}
  \caption{The turn-over semi-major axis for the high inclination Kozai-on case (denoted with ``y" here) v.s. $a_{p0}$ for the  case with $a_B=50$ AU, $e_B=0.3$, $i_B=60^o$, and $M_A=2M_B=M_\odot$.
Planetesimals' orbital eccentricities and inclinations are initially set as very small values ($<10^{-4}$), and their other initial angular orbital elements are set as random values from $0$ to $360^\circ$.
We see that $y\sim a_{p0}$
if $a_{p0}<a_{c}$, i.e., $a_{p0}\lesssim1.5$ AU. For
$1.5\lesssim a_{p0}\lesssim8$ AU, y follows an oscillatory pattern, whose amplitude varies between a lower limit of $a_{c(low)}\sim0.3$ AU and an upper limit of $a_{c}$, i.e., $y\sim0.30-1.5$ AU. For $a_{p0}$ beyond $\sim 8$ AU, the results become chaotic, but the upper and lower limits ($y\sim0.30-1.5$ AU) still hold.
}
    \label{aei2_turn}
   \end{center}
\end{figure}

\begin{figure}
\begin{center}
\includegraphics[width=\textwidth]{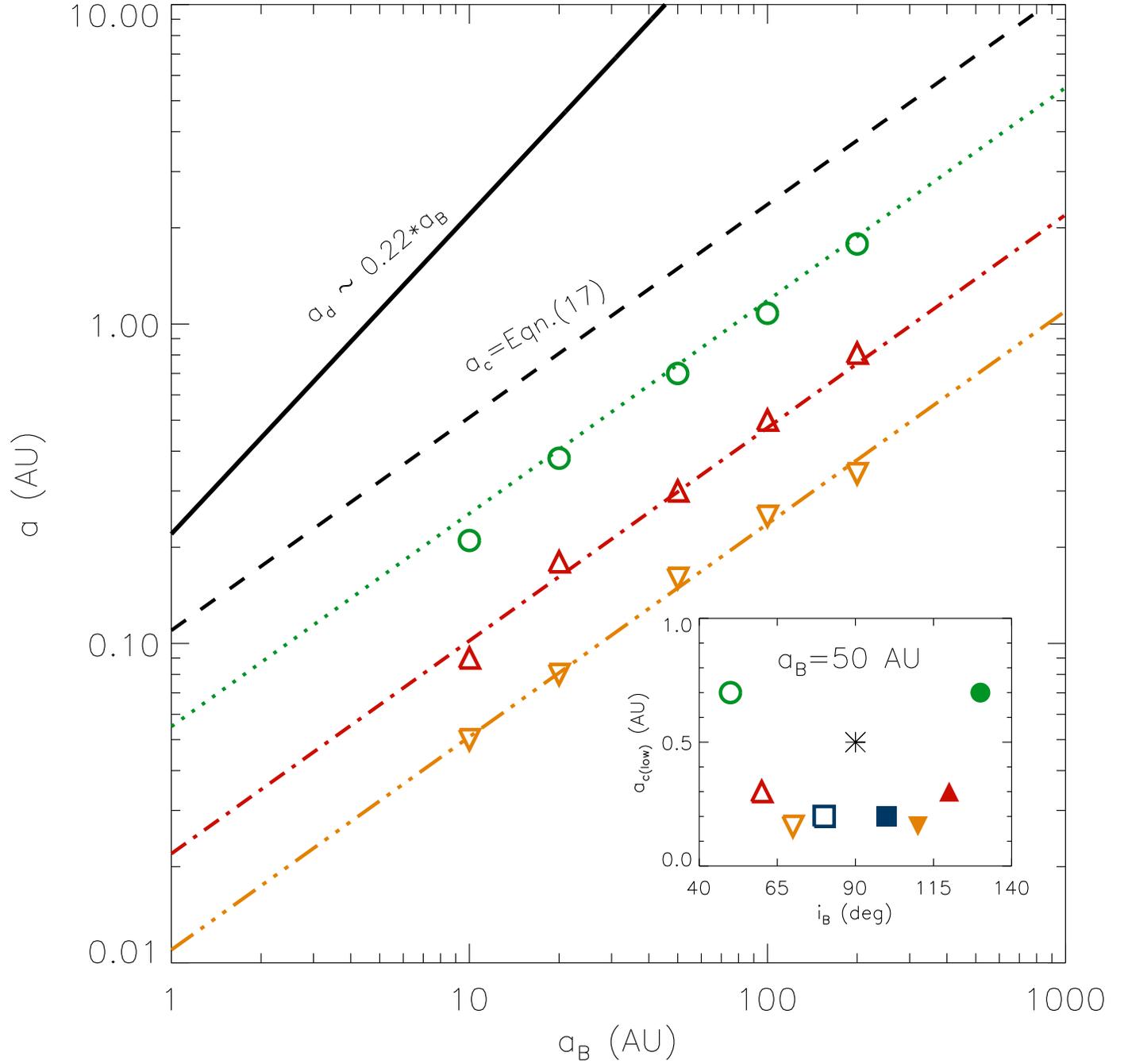}
  \caption{Empirical fitting of $a_{c\,(low)}$
 as a function of $a_B$ for  cases of $i_B=50^\circ$ (dot,  open circle), $60^\circ$
(dashed-dot, open triangle), and $70^\circ$ (dashed-dot-dot, open upside down triangle). The circle, triangle symbols are the results of individual measurements made by
plotting figures similar to Fig.\ref{aei2_turn}.  We also plot the disk stability boundary (solid line), $a_d\sim 0.22 a_B$ for $M_A=2M_B=M_\odot$ and $e_B=0.3$ according to \citet{HW 99} , and $a_{c}$ (dashed) as references.
We find that a good empirical fits for $a_{c(low)}$ can be given as
Eq.\ref{ac_low}. In the inset figure at the bottom right, we also plot the measured
$a_{c(low)}$ as a function of $i_B$. The open symbols are for
prograde cases while the filled ones are for the corresponding
retrograde cases, while the $i=90^\circ$ case is plotted using an
asterisk. We see that the results are symmetrical about
$i=90^\circ$. }
    \label{fac}
   \end{center}
\end{figure}

For the parameters of the standard case, we choose $q=0.5$, $a_B=50$ AU, $e_B=0.3$, $i_B=60^\circ$, and $R_p=5$ km. The results of the standard case (highlighted with thick blue axes) are shown in the second panel of the right-hand column of Fig.\ref{eff_i} and in the middle panels of Fig.\ref{eff_a} and Fig.\ref{eff_qer}, the results of which can be summarized as follows.
\begin{itemize}
\item In the first $2\times10^4$ yr, the planetesimal disk profile varies a lot. In the outer region the profile becomes very noisy with modest fluctuation around $\Sigma_0$. This complicated structure is due to the step-wise jumping as shown in Fig.\ref{aei2} (see the bottom-right panel) and the chaotic behavior near the outer boundary (see Fig.\ref{aei2_turn}). The modest density surge near $\sim2.7$ AU is due to the initial density jump at the ``snowline" $a_{ice}=2.7$ AU. In contrast, in the inner region ($\sim 0.3$ AU) a significant density enhancement (up to $\Sigma/\Sigma_0\sim 10$) is observed, caused by the planetesimal jumping shown in Fig.\ref{aei2} (see the bottom-middle panel) and Fig.\ref{aei2_turn}.
\item At times $\sim10^5$ yr, the planetesimal disk appears truncated near $a_c$ (Eq.\ref{ac}). Within $\sim a_c$, the entire surface density is enhanced by 1 order of magnitude, i.e., $\Sigma/\Sigma_0\sim 10$, while beyond $\sim a_c$, the surface density is entirely depleted down to $\Sigma/\Sigma_0<0.1$. This result is expected from Fig.\ref{aei2_turn} which shows that most, if not all, planetesimals jump into the inner region within $a_c$.
\item At later times ($t\gtrsim10^5$ yr), the planetesimal disk seems to reach a quasi-equilibrium state: the surface density maintains a ``$\daleth$" (daleth) shape profile with the sharp outer boundary from $\sim a_c$ slowly moving inward. This ``orderly" evolution is expected because, as show in Fig.\ref{aei2} (\S\,\ref{on-kozai}), once the planetesimal has completed a jump within $a_c$, its $\hat{e}$ and/or $\hat{i}$ would no longer be excited by the binary perturbations and thus it undergoes slow and smooth inward migration. In other words, the planetesimal's inward migration has transited from B1 regime to SB and finally to the S regime as shown in Fig.\ref{ei} (\S\,\ref{damp_regime}). As the inward migration rate is inverse proportional to its semimajor axis (see \S\,\ref{damp_regime}) in the S regime, the surface density of planetesimals will then decrease from inside to outside on a long timescale (depending on the gas density and the planetesimal size).
\end{itemize}


\subsection{Parameter Exploration}\label{PE}
Through comparisons to the standard case, we will discuss the effects of the modeling parameters, such as $i_B$, $a_B$, $e_B$, $q_B$, $R_p$ and $f_d$.
\subsubsection{Effects of $i_B$}\label{Eff_i}
\begin{figure}
\begin{center}
\includegraphics[width=\textwidth]{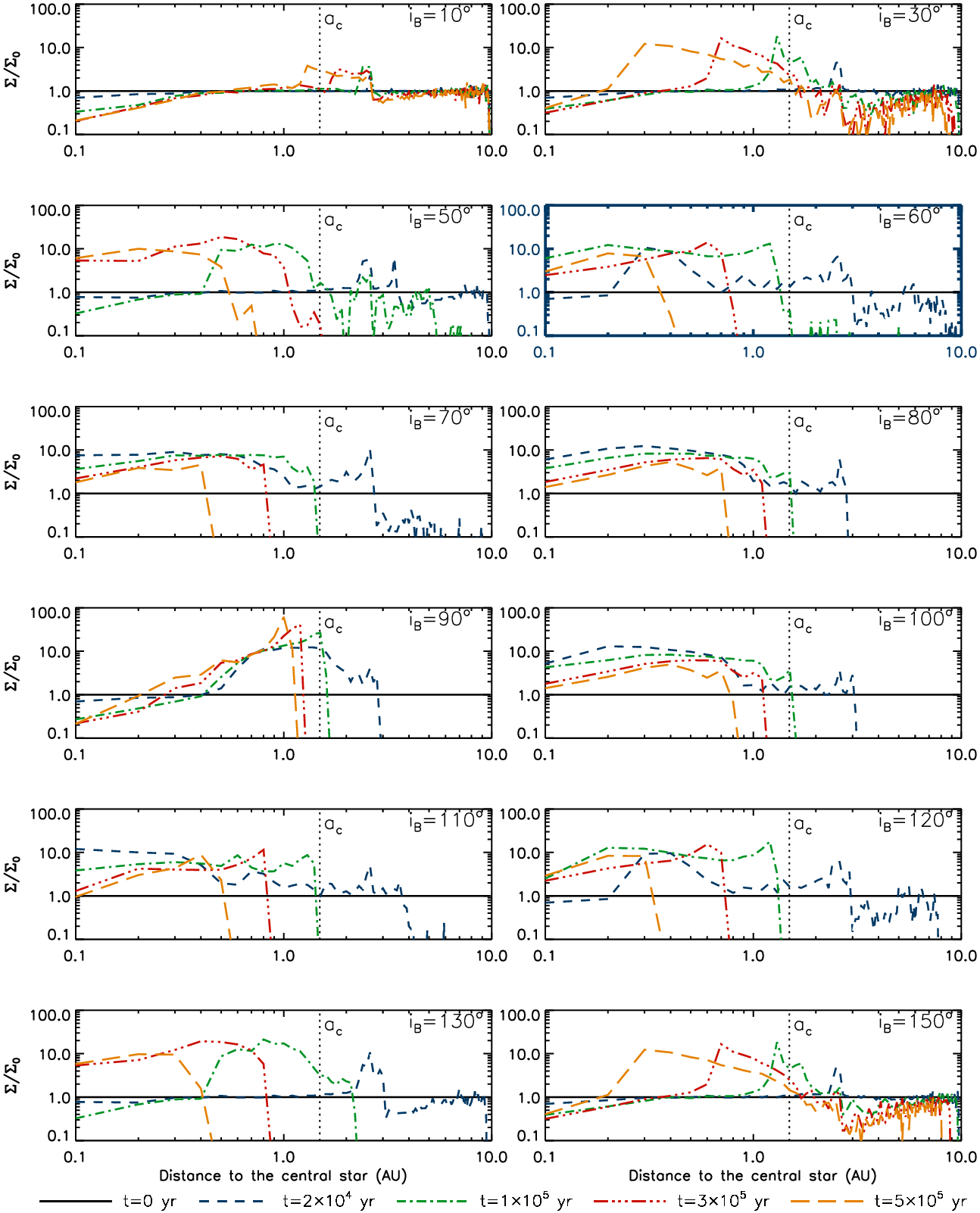}
  \caption{
  Planetesimal evolution for the cases in which $M_A=2M_B=M_\odot$, $a_B=50$ AU, $e_B=0.3$, $i_B=10^\circ, 30^\circ, 50^\circ, 60^\circ,70^\circ,80^\circ$, $R_p=5$ km
  The second panel in the right-hand column (highlighted with blue bold axes) shows the results of the standard case (\S\,\ref{standard}).
  We find in highly inclined cases that the planetesimal disk surface density is highly enhanced ($\Sigma/\Sigma_0\sim10$) in the inner region within $a_c$, while it is significantly depleted beyond $a_c$.  In addition, all results seems to be approximately symmetrical about $i=90^\circ$.  For details see text in \S\,\ref{Eff_i}.
  }

\label{eff_i}
 \end{center}
\end{figure}

\begin{figure}
\begin{center}
\includegraphics[width=0.9\textwidth]{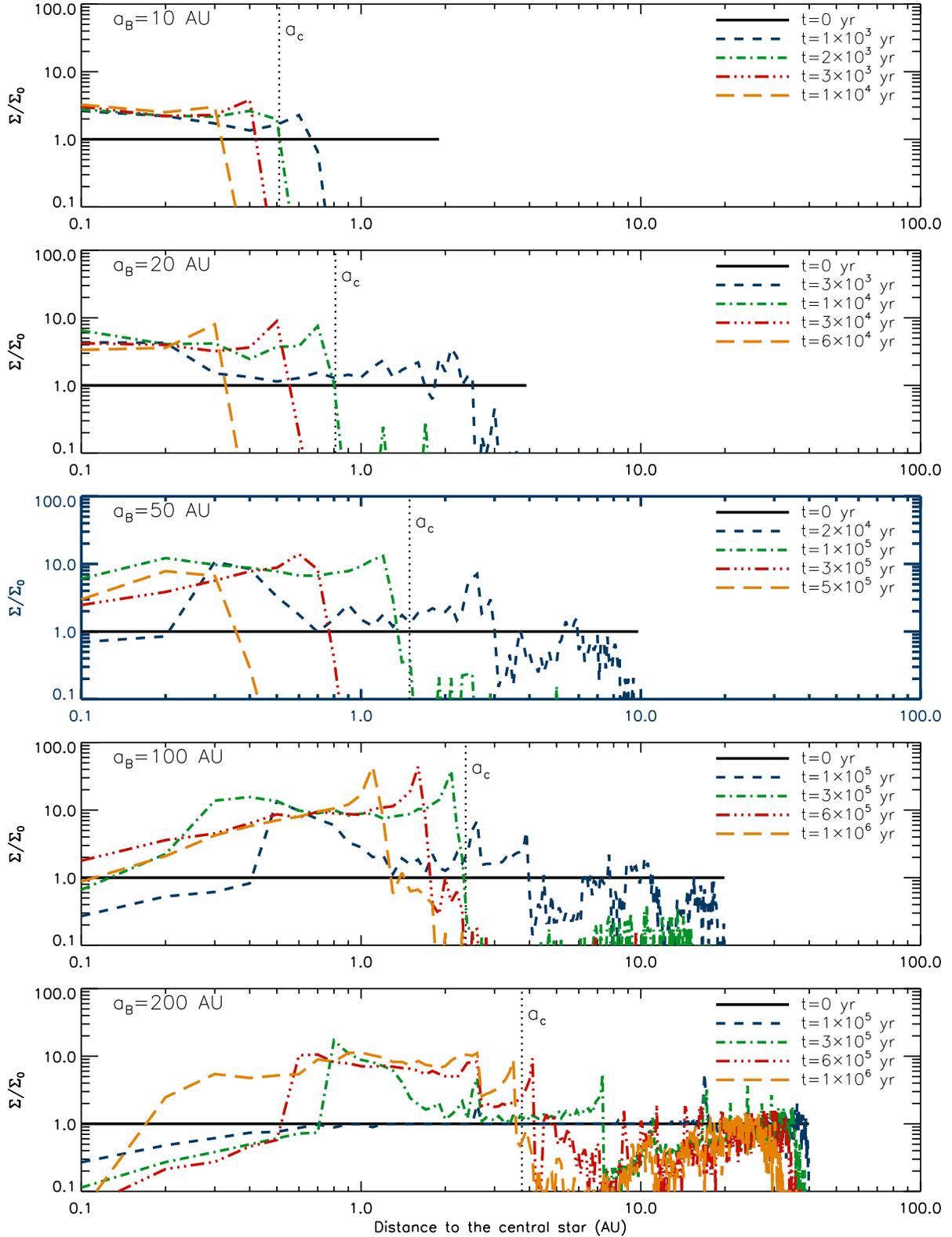}
  \caption{
  Planetesimal evolution for the cases in which $M_A=M_\odot$, $q_B=M_B/M_A=0.5$, $a_B=10, 20, 50, 100, 200, 400$ AU, $e_B=0.3$, $i_B=60^\circ$, $R_p=5$ km.
  The middle panel (highlighted with blue bold axes) shows the results of the standard case (\S\,\ref{standard}).
We see that $a_B$ has significant effects on determining when and to what degree the piling up effects can reach (\S\,\ref{Eff_aB}).  Note, the time-sampling is different in each plot.
  }
\label{eff_a}
\end{center}
\end{figure}

We vary the parameter $i_B$ of the standard case from $10^\circ$ to
$150^\circ$
 but hold all other parameters constant. Our results are plotted in Fig.\ref{eff_i}
(the standard case is in the second panel of the right-hand column of Fig.\ref{eff_i},
outlined in bold), and can be summarized as follows. 
\begin{itemize}
\item Kozai-On
case ($40^\circ <i_B<140^\circ$). Similar to the standard case, the
planetesimal disk reaches a quasi-equilibrium state with a
``$\daleth$" (daleth) profile on a timescale of $\sim10^5$ yr (This
process is a little bit quicker for larger $i_B$ cases). In all
cases, the planetesimals pile up within $a_c$, leading to an average
surface density enhancement up to one order of magnitude, i.e.,
$\Sigma/\Sigma_0\sim 10$. Note, results of the prograde
 cases ($i_B=50^\circ, 60^\circ, 70^\circ, 80^\circ$) and their corresponding retrograde cases ($i_B=130^\circ, 120^\circ, 110^\circ, 120^\circ$) are very similar (see also in Fig.\ref{fac}).
The only slight difference lies in the marginally quicker clearing
of the outer disk.
\item Kozai-Off
case ($i_B=30^\circ, 150^\circ$). In contrast to the Kozai-on case, the planetesimal disk seems to evolve to a unimodal profile on a timescale of $\sim10^5$ yr. This unimodal profile is maintained with a peak enhancement ($\Sigma/\Sigma_0\sim 10$) at $\sim a_c$, slowly moving inward over time. Another feature which distinguishes this from the Kozai-On case is that the disk will not be significantly truncated with respect to the initial disk. Note again, the prograde case ($i_B=30^\circ$) and its corresponding retrograde case ($i_B=150^\circ$) have very similar results.
\item Near-coplanar case ($i_B=10^\circ$). Only modest density enhancement occurs in $\sim$1.5-2.7 AU because planetesimals pass through the ``snowline" ($a_{ice}=2.7$ AU).
\end{itemize}

\subsubsection{Effects of $a_B$}\label{Eff_aB}
We vary the parameter $a_B$ of the standard case from $10$ to $200$ AU but hold all other parameters constant. Results are plotted in Fig.\ref{eff_a} (the standard case is in the middle panel highlighted with blue bold axes), which are summarized as the following.
\begin{itemize}
\item Similar to the standard case, in all cases (except the $a_B=200$ AU case) the planetesimal disks reach a quasi-equilibrium, i.e. a ``$\daleth$" (daleth) shape surface density profile with all planetesimals piling-up in a tight, dense inner region which is truncated near $a_c$. In the $a_B=200$ AU case, the planetesimal disk is not truncated and there are still many planetesimals left beyond 20 AU after $10^6$ yr. This is mainly due to the very low gas density there (thus weak gas drag) and due to the long binary separation (thus weaker perturbations). In addition, the surface density enhancements in the $a_B=10$ and 20 AU cases are significantly lower ($\Sigma/\Sigma_0\sim5$) as compared to the standard case ($a_B = 50$ AU, $\Sigma/\Sigma_0\sim10$)
because the initial planetesimal disks are much smaller (thus less total planetesimals) in such close binary systems.

\item The time for the planetesimal disk to reach a quasi-equilibrium (``$\daleth$-like shape) surface density profile (we define it as $t_{pile}$) is very sensitive to $a_B$. Empirically and to a first order of estimation, $t_{pile}$ is roughly equal to the secular perturbation timescale at $a_{c}$. Substituting the semimajor axis in Eqn.\ref{tsp} with Eqn.\ref{ac}, $t_{pile}$ can be expressed as
\begin{eqnarray}
t_{pile}\sim 28.7\times\left(\frac{a_B}{\rm AU}\right)^2(1-e_B^2) \nonumber \\\left(\frac{M_A}{M_{\odot}}\right)^{1/6}\left(\frac{M_B}{M_{\odot}}\right)^{-2/3} \ \ \ {\rm yr},
\label{tpile}
\end{eqnarray}
which depends mainly on $a_B$ and weakly on $e_B$, $M_A$ and $M_B$. For $a_B>200$ AU, the pile timescale $t_{pile}> 1$ Myr, which is comparable or even longer than the lifetime of the gas disk. Therefore, the jump-piling effects would be very limited for binary systems with $a_B>200$ AU.
\end{itemize}

\subsubsection{Effects of $e_B$, $q_B$, $R_p$ and $f_d$}\label{Eff_other}
\begin{figure}
\begin{center}
\includegraphics[width=\textwidth]{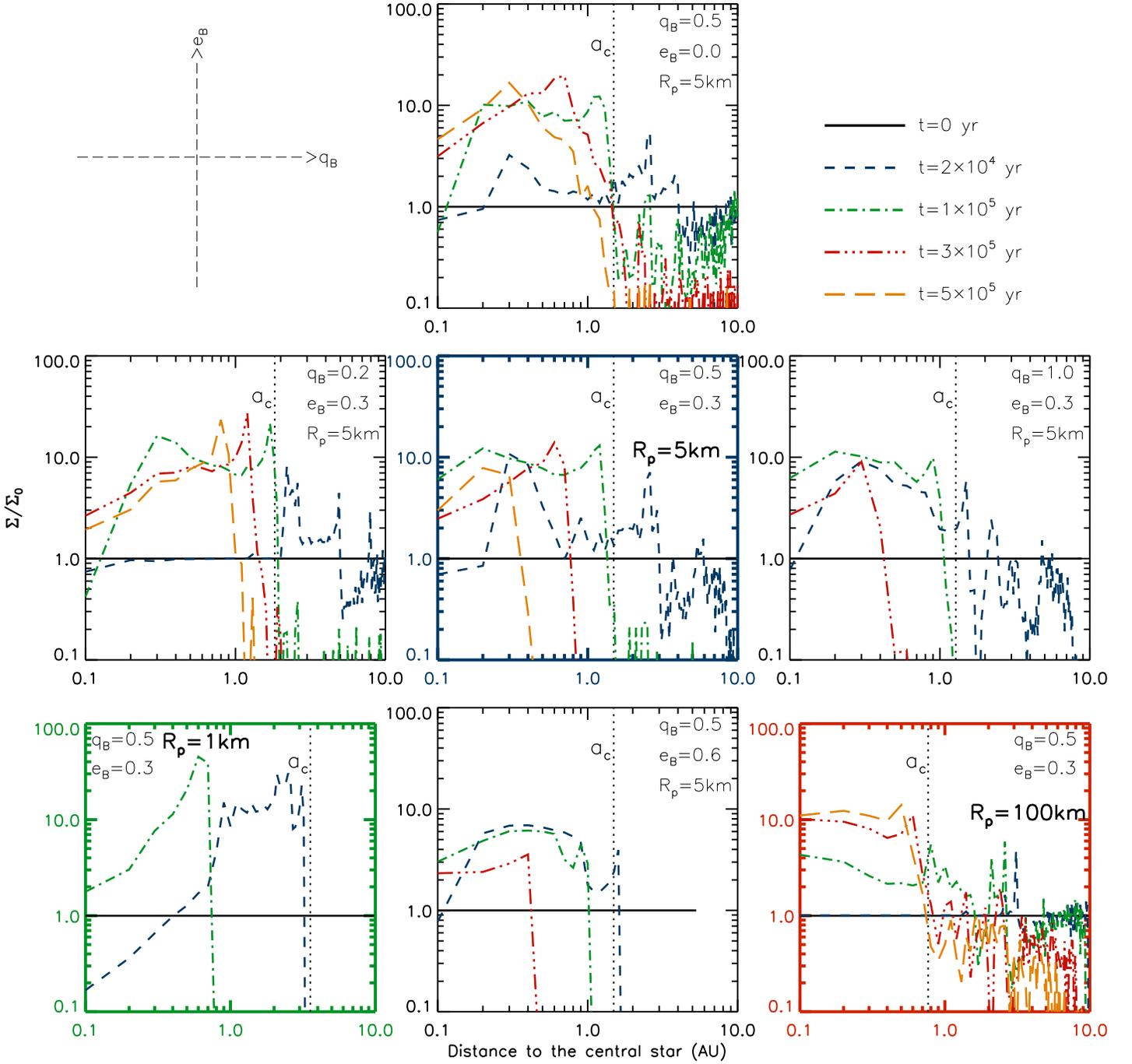}
  \caption{
  Planetesimal evolution for the cases in which $M_A=M_\odot$, $q_B=M_B/M_A=0.2, 0.5, 1.0$, $a_B=50$ AU, $e_B=0, 0.3, 0.6$, $i_B=60^\circ$, $R_p=5, 100$ km.
 The middle panel (highlighted with blue bold axes) shows the results of the standard case (\S\,\ref{standard}).
 The specific effects of $q_B$, $e_B$, $R_p$ and $f_d$ are summarized in \S\,\ref{Eff_other}. In all cases, the general result holds, i.e., disk surface density is highly enhanced in the inner region within $a_c$, while it is significantly depleted beyond $a_c$.
  }
\label{eff_qer}
   \end{center}
\end{figure}

The results are shown in Fig.\ref{eff_qer}, with the standard case located in the middle panel with blue bold axes. We can obtain the effects of $e_B$ and $q_B$ through the comparison of the three middle vertical panels and the comparison of the three horizontal panels, respectively. In addition, we can compare the cases of  $R_p=100$ km and $R_p=1$ km (the two bottom-corner panels) to the standard case (the middle panel with blue bold axes) for the effects of $R_p$. Since increasing the planetesimal radial size ($R_p$) is equivalent to decreasing the gas density($f_d$) from a dynamical view, thus we obtain the effect of $f_d$ at the same time.
\begin{itemize}
\item Effects of $e_B$ ($q_B$). Increasing $e_B$ ($q_B$) leads to quicker evolution of the disk density profile (shorter $t_{pile}$ as also seen the dependence in Eqn.\ref{tpile}) and slightly lower surface density enhancement ($\Sigma/\Sigma_0$) due to the stronger binary perturbation and smaller initial planetesimal disk size.
\item Effects of $R_p$ ($f_d$). Increasing $R_p$ (decreasing $f_d$) is a double-edged sword when it comes to the planetesimal pile-up.
On the one hand, it directly reduces the gas drag acceleration on the planetesimals and thus slows their ``jumping processes" and inhibits the planetesimal piling
(see the leftover material in the outer disk of the $R_p=100$ km case in Fig.\ref{eff_qer}).
On the other hand, once planetesimals have finished their jump into the inner region, they will experience much stronger braking of their further inward migration because of the larger $R_p$ (smaller $f_d$),  which favors the planetesimal pile-up
(see the nearly fixed boundary around $a_c$, and the progressive increase of the surface density within $a_c$ in the $R_p=100$ km case).
The opposite trend can be found by decreasing $R_p$ or increasing $f_d$ (see the case of $R_p=1$ km).

\end{itemize}


\subsection{Analytic Estimate of $\Sigma/\Sigma_0$}
As we have shown above, the inward migration or jumping process
always lead to the formation of a compact, dense disk with an outer
boundary around $a_c$. Therefore, the surface density enhancement in
the inner region can be estimated by assuming that all of the
material initially having $a > a_c$ is transported to $a < a_c$,
leading to an enhancement factor
\begin{eqnarray}
\frac{\Sigma\ }{\Sigma_0}&\sim&\frac{\int_{a_{in}}^{a_{d}}2\pi r\Sigma_0dr}{\int_{a_{in}}^{a_{c}}2\pi r\Sigma_0dr},
\label{ae0}
\end{eqnarray}
where $a_{in}$ (here set as 0.1 AU) is the inner boundary of the planetesimal disk and $a_{d}$ is the initial outer boundary following \citet{HW 99}.  Considering a power law disk profile as $\Sigma_0\propto r^{\alpha}$ ($\alpha\ne2$), then Eq.\ref{ae0} can be rewritten as
\begin{eqnarray}
\frac{\Sigma\ }{\Sigma_0} \sim \left\{ \begin{array}{ll}
\frac{a_{d}^{2+\alpha}-a_{in}^{2+\alpha}}{a_{c}^{2+\alpha}-a_{in}^{2+\alpha}} & \textrm{if  $ a_c<a_{d}<a_{ice}$,\ \ \ \ \ }\\ \\
\frac{a_{ice}^{2+\alpha}-a_{in}^{2+\alpha}+f_{ice}(a_{d}^{2+\alpha}-a_{ice}^{2+\alpha})}{a_{c}^{2+\alpha}-a_{in}^{2+\alpha}} &
\textrm{if  $ a_c<a_{ice}<a_{d}$,\ \ \ \ \ }\\ \\
\frac{a_{ice}^{2+\alpha}-a_{in}^{2+\alpha}+f_{ice}(a_{d}^{2+\alpha}-a_{ice}^{2+\alpha})}{a_{ice}^{2+\alpha}-a_{in}^{2+\alpha}+f_{ice}(a_{c}^{2+\alpha}-a_{ice}^{2+\alpha})} &
\textrm{if  $ a_{ice}<a_c<a_{d}$,\ \ \ \ \ }
\end{array} \right.
\label{ae}
\end{eqnarray}
where $a_{ice}$ is the location of snow line and $f_{ice}$ is the solid density enhancement beyond the snow line. Following \citet{IL 04}, we take $a_{ice}=2.7$ AU and $f_{ice}=4.2$.

\begin{figure}
\begin{center}
\includegraphics[width=\textwidth]{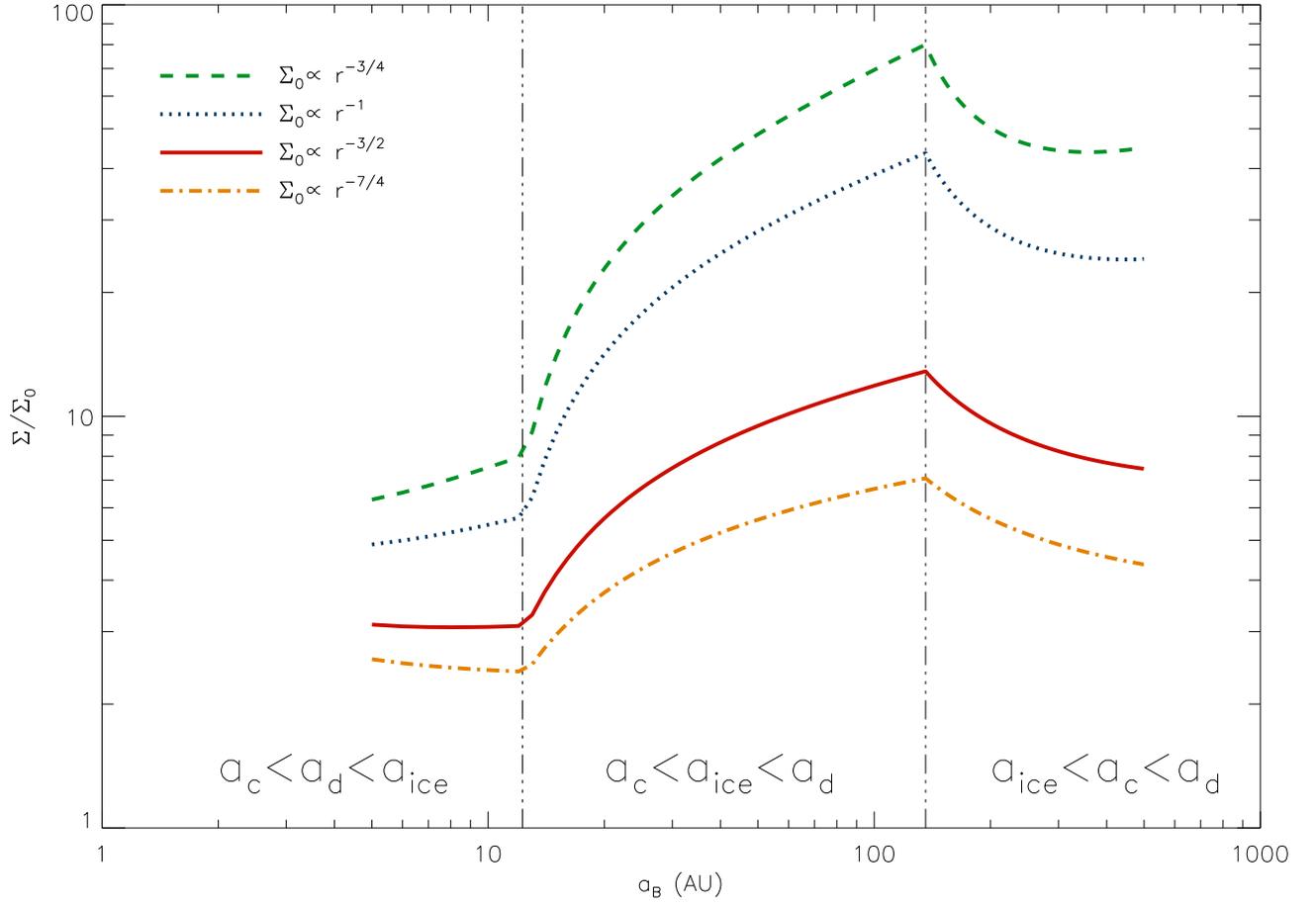}
  \caption{Analytical estimate (Eq.\ref{ae}) of surface density enhancement ($\Sigma/\Sigma$) as a function of semimajor axis of binary systems with fixed orbital eccentricity $e_B=0.3$ and mass ratio $M_B/M_A=0.5$.  Results of four cases with different initial disk profiles are plotted in lines of different styles. The two vertical dash-dots lines, $a_B\sim12.3$ AU and $a_B\sim135.8$ AU mark the turn-over points where $a_d=a_{ice}=2.7$ AU and $a_c=a_{ice}=2.7$ AU, respectively.
 }
\label{sig}
   \end{center}
\end{figure}

Fig.\ref{sig} shows the results of Eq.\ref{ae}. We see that the
analytical estimates are consistent with the N-body simulation
results by comparing the red solid curve of Fig.\ref{sig} to
Fig.\ref{eff_a}. Piling effects are most significant for binary
systems with $a_B\sim100$ AU and for planetesimal disks with flatter
initial disk profiles. While the profile we consider in detail
($\alpha = 3/2$) have surface density enhancements $\sim 10$, for
disk profiles flatter than $\Sigma_0\propto r^{-3/4}$, the surface
density enhancement can be around 2 orders of magnitude.


\section{Discussions}\label{SECN:DISCUSS}
\subsection{Implications for Planet Formation in Highly Inclined Systems}\label{SECN:IMPLICATION}
In this paper, we have shown, for the highly inclined
($i_B\gtrsim30^\circ$) binary systems, that planetesimals from the
outer disk can ``jump'' inwards and pile up in the inner disk. The
typical result of this jumping-piling process is the formation of a
smaller and denser planetesimal disk around the primary, with the
truncation boundary near $a_c$ (Eqn.\ref{ac}), within which the
surface density enhancement is up to $\Sigma/\Sigma_0\sim10$. This
significant change in the planetesimal disk will definitely affect
the subsequent planet-formation processes$-$planetesimal growth, and
the final composition and configuration of the planetary system.

\begin{figure}
\begin{center}
\includegraphics[width=0.85\textwidth]{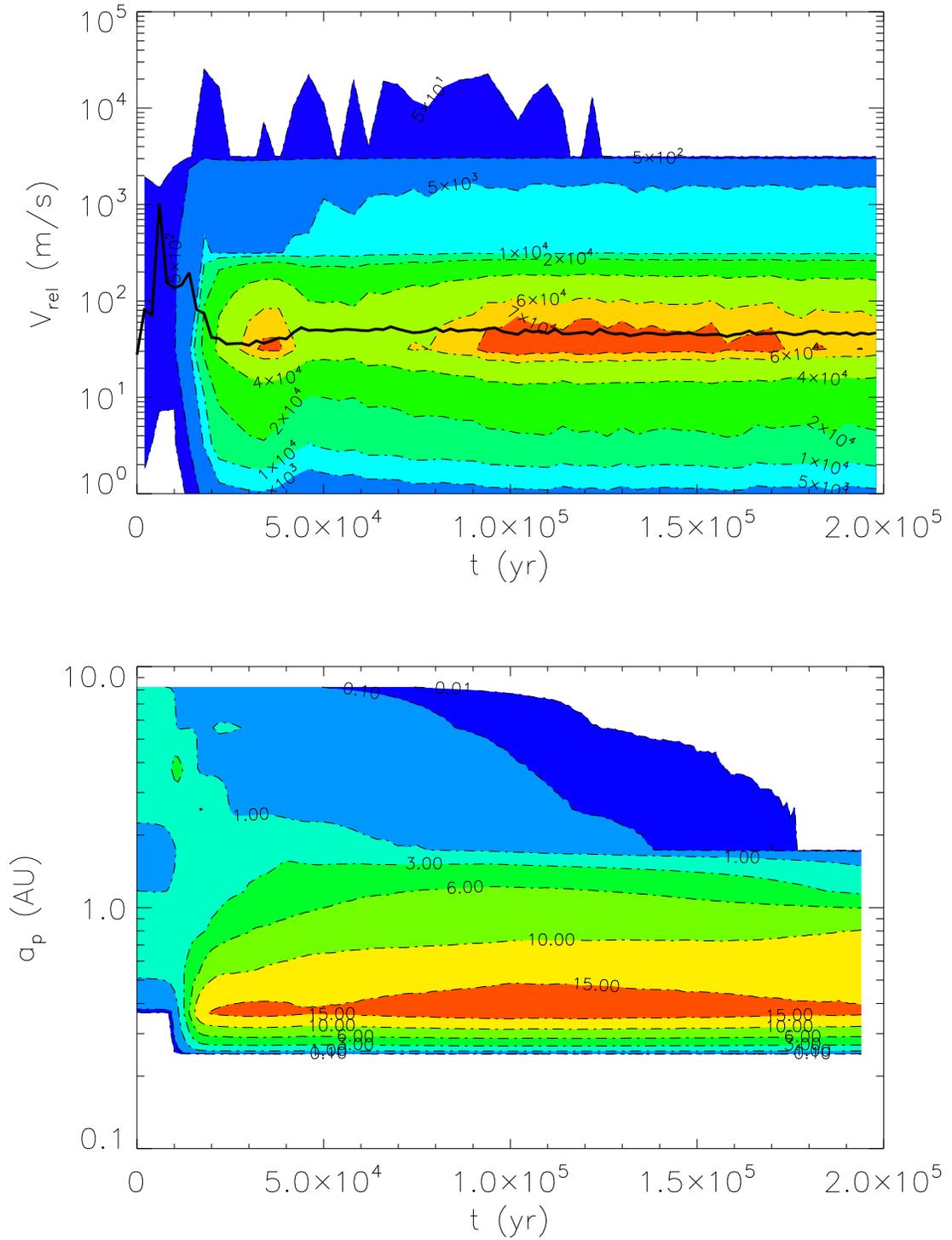}
  \caption{Evolution of relative velocities (top) and semi-major axis (bottom) for 10,000 planetesimals in a system with the standard binary configuration, $M_A=M_\odot$, $q_B=M_B/M_A=0.5$, $a_B=50$ AU, $e_B=0.3$, $i_B=60^\circ$. Planetesimals with a centered Gaussian size distribution between $R_p=1$ km and $R_p=10$ km, are initially randomly distributed between 0.3 AU and 10 AU (flat distribution) with near circular and coplanar orbits. In the top panel, the contours denote the rate of collisions (in arbitrary units), and the thick solid line denotes the evolution of average relative velocity, for all collisions we have searched between 0.3 AU and 10 AU. In the bottom panel, the contours denote the changes in the planetesimal surface density, i.e., $\Sigma_/\Sigma_0$. We see that with the planetesimals piling up in the inner region, their relative velocities and collision rates are significantly reduced and increased respectively, which thus favors planetesimal growth in the inner region.    }
\label{vrel_rate}
   \end{center}
\end{figure}

The two most important parameters which govern planetesimal growth
are, first, the relative velocities ($V_{rel}$) among planetesimals,
and second, their collision rate. The former determines the outcome
of planetesimal-planetesimal collisions, i.e., accretion or erosion;
while the latter controls the speed of planetesimal accretion and/or
erosion.
We investigate in Fig.\ref{vrel_rate} the evolution of these two
parameters during the jumping-piling process for the standard binary
configuration ($M_A=M_\odot$, $q_B=M_B/M_A=0.5$, $a_B=50$ AU,
$e_B=0.3$, $i_B=60^\circ$). In Fig.\ref{vrel_rate}, 10,000
planetesimals with a centered Gaussian size distribution between
$R_p=1$ km and $R_p=10$ km, are initially randomly distributed
between 0.3 AU and 10 AU (flat distribution) with near circular and
coplanar orbits.
As can be seen, in the first $\sim10^4$ yr, the average $V_{rel}$
has a steep increase up to $10^3$ ms$^{-1}$ because of the strong
binary perturbations, and the collision rate is relatively low
because of the large spread of planetesimals at the beginning.
After this point ($t\sim10^4$ yr), planetesimals begin to jump
inwards and pile up in the inner region (within $\sim1.5$ AU),
leading to a surface density enhancement of up to one order of
magnitude, a collision rate enhancement of up to three orders of
magnitude, and a reduction of average $V_{rel}$ down to $40-50$
ms$^{-1}$.
If we adopt a velocity of $50-100$ ms$^{-1}$ as the threshold for
planetesimal erosion as in \citep{Mar 09}, then we see that the
jumping-piling process moves most planetesimals from an erosional
regime to an accretional regime. For this reason, we expect fast
planetesimal growth to occur in the inner region.


In fact, everything (such as $V_{rel}$, $\hat{e}$, and $\hat{i}$ of
planetesimals) within $a_c$, is getting close to that in single star
cases. For example, after planetesimals have completed the jumping
process as shown in Fig.\ref{aei2}, their eccentricities ($\hat{e}$)
are all below $10^{-3}$, which are comparable to that in single star
cases (see Eqn.(10) of \citet{KI 00}) and their inward drifts
transition to the SB or S regime. Therefore, the process of
planetesimal growth within $a_c$ would probably follow a similar way
as in single star systems.
Such a process leads eventually to the formation of planetary embryos (via the runaway and oligarchic growth processes), resulting in a number of embryos which have isolated themselves from one another due to the accretion of all the planetesimals in an annular feeding zone around themselves. This embryo isolation mass is given by \citep{IL 04}
\begin{eqnarray}
M_{\rm iso}\sim 0.16f_d{\rm}^{3/2}f_{\rm ice}^{3/2}\left(\frac{\Sigma}{\Sigma_{\rm 0}}\right)^{3/2}\left(\frac{a}{\rm AU}\right)^{3/4} \ \ \ {\rm M_\oplus}.
\label{Miso}
\end{eqnarray}
where $f_{ice}\sim4.2$ denotes the solid density enhancement beyond
the ice line $a_{ice}=2.7$ AU, and $\Sigma/\Sigma_0$ denotes the
solid surface density enhancement relative to MMSN.

Clearly, in a MMSN, at 1 AU the isolation mass is $0.16 M_\oplus$,
leading to the idea that in the Solar System there may have been a
chain of such $\sim$Mars-mass embryos in the terrestrial planet
region, which subsequently went on to collisionally evolve in a
chaotic growth phase.

However, in the binary scenario we are considering, we have seen
typical surface density enhancements of $\Sigma/\Sigma_0\sim10$ (can
be larger for flatter disk profiles as shown in Fig.\ref{sig}),
leading to an isolation mass at 1 AU $\sim 5\times f_d{\rm}^{3/2} \,
\rm M_\oplus$. Moreover, the initial surface density of the disk
could be above that of the MMSN, so it is entirely plausible to have
$f_d=1-3$, giving isolation masses at 1 AU in the range $5-26 \,\rm
M_\oplus$ and typical separations $10r_H \sim 0.25 - 0.43$ AU.



From the first order estimation above, one could anticipate that in the highly dense inner-disk regions of these inclined binary systems, several giant embryos could form. If these cores grew before the depletion of the gas disk, then several gas-giants could start to form, otherwise, a series of stalled super earths/neptunes may form. In addition, as the planetesimal disk is severely truncated after the jumping-piling process, it is highly probable that such planets would be born in a dynamically compact region with a short stability timescale \citep{Cha 08}. After the depletion of the gas disk, the final configuration of the planetary system would then be further shaped by dynamical evolution mechanisms such as planet-planet scattering \citep{RF 96, For 01, Mal 07}, Kozai migration \citep{WM 03, FT 07, Tak 09} and secular chaos \citep{Mic 06, LW 10, WL 10}.

\subsection{Limitations}\label{SECN:LIMITATION}
First, as presented in \S\,\ref{SECN:MODEL}, we adopt a very simplified gas disk model (CGD) where the gas is in a circular orbit in the circumprimary mid-plane without feeling the companion star as it is in a single star system. In reality, the gas disk structure should be significantly modified by the binary perturbations. Nevertheless, we found that the planetesimal dynamical behaviors are very similar under different models of disk structure for highly inclined ($30^\circ\lesssim i_B\lesssim150^\circ$) binary systems .

Second, the growth and/or erosion of planetesimals due to mutual collisions are not considered in the present paper. Beyond $a_c$, planetesimals are excited onto orbit with large inclinations, leading to their collisional timescale becoming longer than $10^6$ yr (according to Eqn. 12 of Xie et al. 2010), which is larger than the piling timescale $t_{pile}$ (Eqn.\ref{tpile}). Therefore, outer planetesimals can jump and pile up in the inner region before they collide with one another. On the other hand, in the piling region ($<a_c$), planetesimals may grow up quickly if collisions are considered because of the very high collision rate and low relative velocities as shown in Fig.\ref{vrel_rate}. As they grow up, their inward migrations slow down, which can make stronger piling effect in the inner region.

Third, in order to estimate the degree of planetesimal piling effects, we have assumed that \emph{all} planetesimals have already been born before the jumping-piling effect takes place. This assumption is reasonable only if the planetesimals birth rate is very high over the whole disk. However, some theoretical and observational works \citep{CW 06, Sco 07, Nat 07, Cha 10} suggest that such a transition process, from dust to planetesimal, may take as long as a few $10^6$ years, which is comparable to or even larger than the piling timescale $t_{pile}$ (Eqn.\ref{tpile}). In such a case, much smaller, maybe mm-sized, particles piling-up through other mechanisms \citep{YC 04} should be considered.

Finally, perhaps the main uncertainty in the planetesimal ``jumping-piling" effect comes from not including the effect of gas gravity on the planetesimals, i.e. we have neglected disk self-gravity.
One possible effect of gas gravity might be that it can pull highly inclined planetesimals back towards the mid-plane of the gas disk, so that the Kozai effect (and hence the ``jumping-piling" effect) might be inhibited \citep{Fra 11}. However this effect usually happens for massive gas disk. For the standard case we are considering in this paper ($a_B=50$ AU), the disk mass should be larger than $\sim6$ MMSN to switch off the Kozai effect as shown in the Fig.16 of \citet{Fra 11}. For a nominal disk with 1 MMSN, we thus expect that the ``jumping-piling" process will still be efficient even if the disk gravity is included. In fact, a jumping trend of planetesimal semimajor axis can be discerned in Fig.11 of \citet{Fra 11}. In addition, as is effectively shown in the bottom-right panel of Fig.\ref{eff_qer} in our paper, the ``jumping-piling" process is still efficient even for a very tenuous gas disk of less than 0.05 MMSN. Therefore, it is plausible that the planetesimal ``jumping-piling" process described in this paper would be most important during the gas dissipation phase.

\section{Summary}\label{SECN:SUMMARY}
In this paper, we have investigated the intermediate stage of planet formation - from planetesimals to planetary embryos - in highly inclined binary systems with a focus on the effects of gas drag on
planetesimal orbits, especially the evolution in the semimajor axis distribution.

First, we justify our choice of a simplified circular gas disk (CGD) for modeling gas drag force in highly inclined cases. Through numerical tests and comparisons (see Fig.\ref{com_par08} and Fig.\ref{com_4}), we have shown that eccentricity and pericenter precession of the gas disk are not important to the effects of gas drag on a planetesimal for binary systems with inclination $30^\circ\lesssim i_B\lesssim150^\circ$. The major factor in determining the magnitude of the gas drag forces is the mutual inclination $\hat{i}$ between the planetesimal and the gas disk, which is mainly controlled by $i_B$ and is independent of the specific disk model.

Next, we analyze the dynamics of individual planetesimals under the influence of gas drag which leads to the inward migration of plantesimals. Depending on the dominant mechanism which drives the planetesimal eccentricities and inclinations, very different migration behaviors can take place. There are three main regimes to consider (see Fig.\ref{ei}): Regime S, where the situation is close to that in single star systems; Regime B where inward migration is dominated by the binary's perturbations, and an intermediate transition regime SB. In addition, regime B can be divided into three sub-regimes, two of which, the Kozai-off regime and the Kozai-on regime, are studied in detail in the present paper.

For the Kozai-off regime ($i_B<39.2^\circ$ or $i_B>140.8^\circ$), planetesimal inward migration is determined \emph{only} by the excitation and evolution of its orbital inclination ($\hat{i}$). There is a turnover semimajor axis ($a_c$, see Eqn.\ref{ac} and Fig.\ref{aei1}), where planetesimals transit from a fast inward migration (regime B) to a slow one (regime SB or S).  The value of $a_c$ is independent of planetesimal initial semimajor axis ($a_{p0}$), and it can be estimated by equating the timescale of secular perturbation ($t_{sp}$, see Eqn.\ref{tsp}) and the timescale of gas drag damping ($t_{dp}$, see Eqn.\ref{tdp2}).

For the Kozai-on regime ($39.2^\circ<i_B<140.8^\circ$), planetesimals can be excited to orbits with high eccentricities and high inclinations (or even become retrograde with respect to the gas disk). In such a case,  planetesimal inward migration can be very sudden. Planetesimals in the outer disk can directly jump into the inner disk where they then slow their inward migration as their orbits  circularize and become coplanar due to strong gas drag damping (see Fig.\ref{aei2}). Unlike the Kozai-off case, the turnover semimajor axis for this migration transition is no longer independent of $a_{p0}$, but it is always located in a range between $a_{c(low)}$ (see, Fig.\ref{aei2_turn} and Fig.\ref{fac}) and $a_c$.

We then studied the behavior of a swarm of planetesimals for both Kozai-on and Kozai-off cases with an exploration of many parameters (see Fig.\ref{eff_i}, Fig.\ref{eff_a}, and Fig.\ref{eff_qer}), including the binary mass ratio ($q_B$), semimajor axis ($a_B$), eccentricity ($e_B$), inclination ($i_B$), gas density scale ($f_d$) and planetesimal radial size ($R_p$). A robust  result is that planetesimals migrate/jump inwards and pile up in the inner region, leading to a smaller and denser planetesimal disk with a truncatation boundary near $a_c$ and a surface density enhancement or order $\Sigma/\Sigma_0\sim10$. The timescale for such a ``jumping-piling" process to operate, $t_{pile}$ (see Eqn.\ref{tpile}), is mainly dependent on $a_B$.

Applying such ``jumping-piling" effects to planet formation, we expect a growth-favored region for planetesimals in the inner disk (within $a_c$), where collision rates are high and relative velocities are low (see Fig.\ref{vrel_rate}). Such conditions \emph{may} lead to the formation of embryos sufficiently massive to undergo runaway gas accretion (provided they form prior to the dissipation of the gas disk).

This work, focusing only on the effect of gas drag on a planetesimal, is our first step towards understanding planet formation in highly inclined binary systems. Future work needs to account for important physical factors such as the effects of gas disk gravity, planetesimal accretion/fragmentation, and the time at which planetesimals emerge and the jump-piling process begins.


\section{Acknowledgments}
 This work was supported by the National Natural Science Foundation of China
  (Nos.10833001, 10778603 and 10925313), the National Basic Research Program of China(No.2007CB814800), NSF with grant AST-0705139 and 0707203, NASA with grant NNX07AP14G and NNX08AR04G, W.M. Keck Foundation, Nanjing University and also University of Florida. J.-W. Xie was also supported by the China Scholarship Council and China Ministry of Education. M. J. Payne was also supported by the NASA Origins of Solar Systems grant NNX09AB35G and the University of Florida's College of Liberal Arts and Sciences.


%
\clearpage

\clearpage

\clearpage


\clearpage

\clearpage


\clearpage

\clearpage


\clearpage

\clearpage

\clearpage

\clearpage

\clearpage

\clearpage

%

\clearpage
\appendix

\section{maximum forced eccentricity}
If the secular perturbation (Kozai effect off) dominates the orbital evolution of a planetesimal. Then the average eccentricity of the planetesimal can be estimated as its forced eccentricity $e_f$, where
\begin{equation}
e_{f}\sim\frac{5}{4}\frac{a}{a_B}e_B,
\label{ef}
\end{equation}
which is proportion to the planetesimal semimajor axis. Thus the maximum $e_f$ is obtained at the edge of the circumprimary disk. According to \citet{Pic 05}, the radial size of the circumprimary  disk can be estimated as
\begin{equation}
R_d\sim R_{Egg}0.733(1-e_B)^{1.20}q^{0.07}
\end{equation},
where $R_{Egg}$ is the radius of the Roche lobe calculated by Eggleton (1983),
\begin{equation}
R_{Egg}=\frac{0.49q_1}{0.6q_1^{2/3}+{\rm ln}(1+q_1^{1/3})}a_B.
\end{equation}

Here, $q_1=M_A/M_B$ and $q2=M_B/(M_A+M_B)$ are two mass ratios, and $M_A$ and $M_B$ are the masses of the primary and the secondary respectively. For a typical binary mass ratio, $q_1=1/2$, we have $R_d\sim0.38a_B(1-e_B)^{1.2}$. Substituting this $R_d$ into Eq.\ref{ef}, then the maximum forced eccentricity is $e_f\sim0.1$.


%
%

\end{document}